\documentclass[a4paper,11pt]{article}
\pdfoutput=1 
\usepackage{jheppub} 
\usepackage{graphicx} 
\usepackage{amsmath,amsfonts}  
\usepackage{hyperref} 
\usepackage{cleveref}
\usepackage{mathrsfs}
\usepackage{subfigure}
\usepackage[dvipsnames]{xcolor}
\usepackage{comment}
\usepackage[normalem]{ulem}
\usepackage{amsmath}
\usepackage{placeins}

 \crefname{section}{Sec.}{Secs.}
 \crefname{equation}{Eq.}{Eqs.}
 \crefname{figure}{Fig.}{Figs.}
 \crefname{table}{Tab.}{Tabs.}
 \crefname{appendix}{App.}{Apps.}

\newcommand{\gt}{{\tilde{g}}}

\newcommand{\A}{a}
\newcommand{\B}{b}

\newcommand{\At}{\tilde{a}}
\newcommand{\Bt}{\tilde{b}}

\newcommand{\mn}{{\mu \nu}}
\newcommand{\et}{\tilde{\varepsilon}}
\newcommand{\pt}{\tilde{p}}
\newcommand{\vt}{\tilde{v}}
\newcommand{\pit}{\tilde{\pi}}

\newcommand{\etat}{\tilde{\eta}}
\newcommand{\taut}{\tilde{\tau}}
\newcommand{\Pit}{\tilde{\Pi}}
\newcommand{\sigt}{\tilde{\sigma}}
\newcommand{\Ut}{\tilde{u}}
\newcommand{\nabt}{\tilde{\nabla}}
\newcommand{\Ctaut}{\tilde{C}_\tau}
\newcommand{\Cet}{\tilde{C}_\eta}

\newcommand{\freq}{\mathfrak{w}}
\newcommand{\momentum}{\mathfrak{q}}

\newcommand{\D}{\delta}

\newcommand{\nt}{n_2}

\newcommand{\Mbh}{M}

\newcommand{\imag}{\mathrm{i}}

\newcommand{\gtx}{g_{tx}}
\newcommand{\gyx}{g_{yx}}
\newcommand{\gtt}{g_{tt}}
\newcommand{\gxx}{g_{xx}}
\newcommand{\gyy}{g_{yy}}
\newcommand{\gty}{g_{ty}}
\newcommand{\gttx}{\tilde{g}_{tx}}
\newcommand{\gtyx}{\tilde{g}_{yx}}
\newcommand{\gttt}{\tilde{g}_{tt}}
\newcommand{\gtxx}{\tilde{g}_{xx}}
\newcommand{\gtyy}{\tilde{g}_{yy}}
\newcommand{\gtty}{\tilde{g}_{ty}}
\newcommand{\zshn}{z_{sh}^{(3)}}
\newcommand{\zson}{z_{so}^{(3)}}

\newcommand{\Zsh}{Z_{sh}}
\newcommand{\Zso}{Z_{so}}
\newcommand{\Gtx}{G_{tx}}
\newcommand{\Gyx}{G_{yx}}
\newcommand{\Gtt}{G_{tt}}
\newcommand{\Gxx}{G_{xx}}
\newcommand{\Gyy}{G_{yy}}
\newcommand{\Gty}{G_{ty}}
\newcommand{\gyxn}{g^{(3)}_{yx}}
\newcommand{\gtxn}{g_{tx}^{(3)}}
\newcommand{\gxxn}{g_{xx}^{(3)}}
\newcommand{\gyyn}{g_{yy}^{(3)}}
\newcommand{\gtyn}{g_{ty}^{(3)}}
\newcommand{\gttn}{g_{tt}^{(3)}}
\newcommand{\gttres}{g_{tt}^{(1)}}

\newcommand{\pityx}{\tilde{\pi}_{yx}}

\newcommand{\pitxx}{\tilde{\pi}_{xx}}

\begin{document}

\title{On prethermal time crystals from semi-holography \\
}

\author[a]{Toshali Mitra,}
\author[b]{Sukrut Mondkar,}
\author[c]{Ayan Mukhopadhyay,}
\author[d]{and Alexander Soloviev}

\affiliation[a]{Institute for Theoretical Physics, University of Heidelberg, D-69120 Heidelberg, Germany}
\affiliation[b]{Harish-Chandra Research Institute, A CI of Homi Bhabha National Institute, Chhatnag Road, Jhunsi, Prayagraj (Allahabad) 211019, India}
\affiliation[c]{Instituto de F\'{\i}sica, Pontificia Universidad Cat\'{o}lica de Valpara\'{\i}so,
Avenida Universidad 330, Valpara\'{\i}so, Chile.}
\affiliation[d]{Faculty of Mathematics and Physics, University of Ljubljana, Jadranska ulica 19, SI-1000, Ljubljana, Slovenia}

\emailAdd{t.mitra@thphys.uni-heidelberg.de}
\emailAdd{sukrutmondkar@hri.res.in}
\emailAdd{ayan.mukhopadhyay@pucv.cl}
\emailAdd{alexander.soloviev@fmf.uni-lj.si}

\abstract{We demonstrate the existence of a pair of almost dissipationless oscillating modes at low temperatures in both the shear and sound channels of a hybrid quantum system, comprised of a weakly self-interacting perturbative sector coupled to strongly self-interacting holographic degrees of freedom described by a black hole geometry. We argue that these modes realize prethermal time-crystal behavior in semi-holographic systems without fine-tuning and can be observed by measuring operators that probe either the hard (perturbative) or the soft (holographic) sector. We also find novel {short wavelength} instabilities that lead to the formation of inhomogeneities even at higher temperatures.
These results provide evidence that black holes with planar horizons and dynamical boundary conditions can develop both inhomogeneous and metastable time-crystal phases over a wide range of temperatures set by an intermediate scale given by the intersector coupling. Furthermore, they suggest that such phases can be realized without external driving in non-Abelian plasmas of asymptotically free gauge theories in the large-$N$ limit.}

\maketitle

\section{Introduction}\label{sec:intro}
The gauge/gravity duality (or holography) provides a correspondence between certain quantum gauge theories and higher-dimensional gravitational systems~\cite{Maldacena:1997re,Ammon:2015wua}. Because the duality is intrinsically strong–weak, mapping strongly coupled dynamics in one description to weakly coupled physics in the other, it has become a powerful framework for analyzing strongly coupled quantum many-body phenomena. Semi-holography~\cite{Faulkner:2010tq, Mukhopadhyay:2013dqa,Iancu:2014ava} extends this idea by self-consistently coupling a weakly interacting sector to a strongly interacting holographic one. This hybrid construction has proved promising for systems where strongly and weakly coupled degrees of freedom coexist, with applications ranging from the quark–gluon plasma (QGP)~\cite{Iancu:2014ava,Mukhopadhyay:2015smb,Mukhopadhyay:2016fkl,Banerjee:2017ozx,Kurkela:2018dku,Ecker:2018ucc,Mitra:2020mei,Mitra:2022xtb,Mitra:2025afp} to non-Fermi liquids~\cite{Mukhopadhyay:2013dqa,Doucot:2017bdm,Doucot:2020fvy,Samanta:2022myh,Doucot:2024hzq}, and more recently to the dynamics of symmetry breaking in hybrid quantum systems~\cite{Mondkar:2021qsf} (see also \cite{Kibe:2023ixa,Mondkar:2023gld}). 

In the following, we briefly review time-crystal phases, argue why such dynamical phases can be realized in non-Abelian plasmas of asymptotically free gauge theories generically using the large-$N$ semi-holographic description, and summarize our results.

\paragraph{Prethermal time crystals:} Time crystals are states of matter breaking time translation symmetry \cite{Wilczek:2012jt, PhysRevLett.109.160402,Sacha_2018, Yao:2018zzv, Khemani:2019nzi, time-crystal-book-sacha, Guo_2020, annurev:10.1146, 10.1088/978-0-7503-3563-8ch6, Hannaford_2022, RevModPhys.95.031001}. Although such states cannot exist in thermal equilibrium \cite{Bruno:2012ddy, Bruno:2013mva, Nozieres_2013, Watanabe:2014hea}, these can be realized away from equilibrium in an open/closed system with driving \cite{PhysRevLett.117.090402, Sacha_2018, Khemani:2019nzi, annurev:10.1146, RevModPhys.95.031001}. In the case of external Floquet driving, the time crystal state breaks the discrete time-translation symmetry of the drive \textit{spontaneously} demonstrating robust sub-harmonic oscillations irrespective of initial conditions and can be produced by a mechanism such as \textit{prethermalization} that prevents heating of the system (under continuous driving) to infinite temperature\footnote{Another proposed mechanism is many-body localization~\cite{Zhang:2016kpq, PhysRevLett.117.090402, PhysRevLett.116.250401,  PhysRevB.94.085112, PhysRevB.94.085112, PhysRevLett.119.010602, annurev:10.1146, RevModPhys.95.031001, Randall:2021ggc}, which requires both strong disorder and short-range interactions and is therefore difficult to realize \cite{Sacha_2018, Khemani:2019nzi, PhysRevB.95.155129, PhysRevB.99.205149, PhysRevLett.119.010602, RevModPhys.95.031001}.}\cite{Else:2016ags, Kuwahara:2016lfr, Abanin_2017, PhysRevB.95.014112, PhysRevLett.123.210602}. Prethermalization is realized if the system has an intrinsic energy scale $J$, is driven at a frequency $\omega_d \gg J$, and  
makes
$\omega_d/J$ local arrangements to absorb a quantum $\omega$ frequency from the drive. As a result, the heating timescale is $\tau_*\approx \exp(\omega_d/J)$. In this case, if the full system also has a $\mathbb{Z}_N$ symmetry, then it can be described by an effective local static Hamiltonian $D_*$, such that the unitary time evolution over the driving period $T = 2\pi/\omega_d$ is 
\[
U_f\approx X e^{-iTD_*}, \,\,{\rm with}\,\, X^N =1,\,\,{\rm and}\,\, [X, D_*] =0,
\]
where $X$ is the $\mathbb{Z}_N$ symmetry transformation. The time-crystal behavior characterized by oscillations over time periods $NT$ is realized if the state $\rho$ \textit{prethermalizes}, implying that $\rho\propto e^{-\beta D_*}$ but breaks the emergent $\mathbb{Z}_N$ symmetry spontaneously, i.e. $\rho \neq X \rho X^\dagger$. Clearly, such a mechanism can be realized only at low temperatures required for the spontaneous breaking of the $\mathbb{Z}_N$ symmetry. 
It has been shown that the prethermal time-crystal behavior can be realized even at high temperatures when an approximately conserved symmetry, e.g., $U(1)$ as discussed in \cite{Luitz:2020ndr}, emerges. The experimental observation of such a prethermal discrete $U(1)$ time crystal has been reported in \cite{Kyprianidis:2021onp,Stasiuk:2023tsp}.

\paragraph{Semi-holographic setups:}\label{sec:intro::par:semi-holo} The semi-holographic approach provides a consistent description of processes in which \textit{both} perturbative ultraviolet (hard) and non-perturbative infrared (soft) degrees of freedom play a significant role \cite{Faulkner:2010tq, Mukhopadhyay:2013dqa,Iancu:2014ava, Mukhopadhyay:2015smb, Mukhopadhyay:2016fkl, Banerjee:2017ozx, Doucot:2017bdm,Kurkela:2018dku, Ecker:2018ucc, Mitra:2020mei, Doucot:2020fvy, Mondkar:2021qsf, Mitra:2022xtb, Samanta:2022myh, Doucot:2024hzq, Mitra:2025afp}. It is assumed that the non-perturbative degrees of freedom are strongly coupled and admit a holographic dual description in terms of a classical gravitational theory in one higher dimension in the large-$N$ limit. In order to be consistent with an effective Wilsonian description at any energy scale, it is necessary that operators such as the energy-momentum tensor of the full system can be constructed from the effective degrees of freedom of both hard and soft subsystems without requiring further microscopic knowledge beyond the specified scale of coarse-graining. This necessitates a special form of coupling called \textit{democratic coupling}~\cite{Banerjee:2017ozx,Kurkela:2018dku} between the two subsystems in which the sources of the marginal/relevant operators of the subsystems are determined self-consistently via local algebraic equations in terms of the state-dependent expectation values of the operators and the actual physical (vanishing/non-vanishing) source. This allows for the explicit construction of the full energy-momentum tensor of the system at any level of coarse-graining that is locally conserved in the actual physical background metric (which is the Minkowski metric for practical purposes). The full system with democratic couplings can be described via an action formalism as summarized in   \cref{sec:setup::subsec:action-formalism} below. The scale of the intersystem democratic coupling is state-dependent, e.g., in the case of heavy-ion collisions, it is the saturation scale \cite{Iancu:2014ava}.

\paragraph{Time-crystal behavior in large-$N$ semi-holography:} In this work, we will establish that the semi-holographic description provides a natural setup in which time-crystal behavior can arise in the large-$N$ limit via the mechanism of prethermalization. In the large-$N$ limit, each subsystem has an entropy current, and the full system's entropy current is the sum of these subsystem entropy currents. As a result, the full isolated system can reach a pseudo-equilibrium state in which two subsystems have different physical temperatures, raising a puzzle as to whether thermalization of the full system is realized in the large-$N$ limit. This issue has been resolved by the authors in \cite{Mitra:2025afp} recently by showing that (i) generally the global equilibrium state in which both subsystems have the same physical temperature has the maximum entropy in the microcanonical ensemble (not obvious a priori in the democratic coupling scheme), and (ii) the ratio of the difference of the physical temperatures of the two subsystems of the full isolated system to the temperature of the global equilibrium at the same total energy fall as a power of the average energy density for typical initial conditions (see \cite{Mitra:2025afp} for a more precise statement).

Therefore, the full system is typically very close to actual thermal equilibrium in the large-$N$ limit for finite and large average energy densities, and the actual thermal equilibration (in which the tiny difference of the subsystem temperatures subsides) takes place over an $\mathcal{O}(N^2)$ timescale. This implies that the hard degrees of freedom prethermalize in the background of the strongly interacting soft modes reminiscent of bottom-up thermalization \cite{Baier:2000sb}\footnote{See also \cite{Mitra:2022xtb} for semi-holographic hydrodynamization in the context of heavy-ion collisions.}. During this process, the entropy current of each subsystem is approximately conserved and can be thought of as emergent local conservation laws. Therefore, time-crystal behavior can be realized if the full system has an almost dissipationless inhomogeneous mode with finite frequency that is constituted primarily by the hard (or soft) sector.\footnote{Although prethermal fixed points can be seen in effective kinetic descriptions of gauge theories \cite{Berges:2004ce,Berges:2020fwq}, there is no good argument why time-crystal-like behavior can arise here.} The time-crystal behavior can be observed by measuring operators that probe the hard (or soft) sector in the appropriate channel. Crucially, the time-crystal behavior can be possibly realized irrespective of initial conditions without any explicit external driving. 

\paragraph{Summary of results:} We model the hard sector via M\"uller-Israel–Stewart (MIS) description \cite{mueller,israel} and study the collective modes of the full semi-holographic system in which the hard sector is coupled to the holographic soft sector via the effective metric coupling~\cite{Kurkela:2018dku}. For reasons described above, we can assume that the full system has thermalized to a good degree of approximation (the results do not change if the two subsystems have tiny differences in temperatures). The soft sector is described by a black hole with a planar horizon. The dissipationless inhomogeneous prethermal time crystal (PTC) mode with finite frequency arises from the modification of the k-gap phenomenon of the weakly self-interacting hard sector in the shear channel. The k-gap phenomenon is the merging of the relaxation mode and the shear hydrodynamic mode at a finite momentum observed generally in liquids \cite{PhysRevB.101.214312} which is naturally explained by the MIS description (see \cite{Baggioli:2019jcm} for an excellent discussion). In semi-holography, this k-gap phenomenon is modified by the influence of the soft sector in which a pair of dissipationless modes with finite frequency appears at a higher value of the momentum $q_g$ for sufficiently low temperatures. We show that our calculations can be explained by a simple quasi-hydrodynamic theory \cite{Grozdanov:2018fic, Baggioli:2023tlc} of the hard sector, which mimics the democratic effective coupling of semi-holography. As expected, $q_g$ diverges when the intersystem coupling goes to zero, but can be reasonably small at higher temperatures/intersystem coupling. Crucially, the PTC behavior appears in the shear channel at low temperatures without any fine-tuning. 

Interestingly, we also find that almost dissipationless PTC-type modes with finite momentum also appear in the soft sector in the \textit{sound channel} by the modification of the black hole quasinormal modes due to the influence of the hard sector at low temperatures. Although this can be attributed to the nature of the coupling, it is harder to describe this behavior in terms of a simplified quasi-hydrodynamic theory or to study this numerically.

{Finally, we find inhomogeneous zero modes (IZM) leading to short wavelength instability in the semi-holographic sound channel. The hybrid mode originating from the relaxation mode of the hard MIS sector becomes gapless exactly at a finite momentum $q_0 \neq 0$ signaling an IZM and an emergent spatial scale in the semi-holographic system. These IZMs occur at all temperatures up to the cutoff of the model. Unlike the Gregory-Laflamme (GL) instability of higher-dimensional black strings~\cite{Gregory:1993vy, Gregory:1994bj, Gregory:2011kh, Lehner:2011wc} which occurs at long wavelengths, we find that modes with momenta $q> q_0$ are unstable. This feature of the instability of high momentum modes carrying momentum larger than that of an IZM has been found earlier in the semi-holographic setup of \cite{Mondkar:2021qsf}. To our knowledge, such short wavelength instabilities are novel. We show that such IZMs can also be obtained from a simpler quasi-hydrodynamic theory of the hard sector which mimics the inter-system coupling.}

Our analysis thus uncovers possible new dynamical phases of black holes with planar horizons without fine-tuning. At low temperatures, the coupled system supports long-lived oscillatory excitations resembling prethermal time crystals; at all temperatures, it admits finite-momentum zero modes that drive inhomogeneous phases; and at higher momenta, the system becomes {unstable despite having only two-derivative Einstein dynamics in the bulk.} 
These phenomena show that semi-holography enables qualitatively new phases of black-brane dynamics and hybrid quantum systems. To establish these features firmly, we would require solving the full non-linear equations, which we leave to future work.

The plan of the paper is as follows: we review the semi-holographic formulation in Sec. \ref{sec:setup} and describe the methodology of computing hybrid quasinormal modes in Sec. \ref{sec:methodology}. In Sec. \ref{sec:results}, we present the results in both shear and sound channels with special emphasis on the modes which lead to pre-thermal time crystal behavior and inhomogeneous ground states. Finally, we conclude in Sec. \ref{sec:discussion} with discussion and outlook.

\section{The semi-holographic framework}\label{sec:setup}
\subsection{Action formalism for semi-holography}\label{sec:setup::subsec:action-formalism}

Semi-holography provides a framework in which a weakly coupled (perturbative) sector and a strongly coupled (holographic) sector are coupled in a way that remains consistent with the Wilsonian renormalization group (RG) at all energy scales. In particular, the so-called democratic couplings can be systematically derived from a semi-holographic Wilsonian effective action \cite{Banerjee:2017ozx,Kurkela:2018dku}.
The two subsectors do not exchange degrees of freedom directly; instead, each is deformed by promoting its marginal and relevant couplings to local algebraic functionals of operators of the two sectors. 
The resulting marginal/relevant deformations of both subsectors render the full effective action renormalizable, while the underlying action principle guarantees the existence of a total energy-momentum tensor which is conserved in the physical background metric. In general, however, each subsector propagates in a distinct effective metric, reflecting its state-dependent marginal deformations.

An important advantage of this construction is that it naturally lends itself to phenomenology. It can be shown that the energy-momentum tensor of the full system (conserved in the physical background metric) can be written as a local polynomial in the energy-momentum tensors of the individual perturbative and non-perturbative (holographic) subsectors \cite{Banerjee:2017ozx,Kurkela:2018dku}. Consequently, once coarse-grained descriptions of the perturbative and non-perturbative (holographic) subsectors are known at any scale, they are sufficient to build a coarse-grained description of the full hybrid system at the same scale in consistency with Wilsonian RG. This feature is special to semi-holographic (democratic) couplings and would not hold for arbitrary interactions between the perturbative and non-perturbative subsectors. For instance, consider a mixture of water and oil. Knowing only the hydrodynamic transport data of water and of oil is not enough to determine the hydrodynamics of the mixture because the transport coefficients of the mixture also depend on the detailed intermolecular interactions. By contrast, semi-holographic democratic couplings allow us to construct the hydrodynamic description of the full gauge theory from the coarse-grained hydrodynamic properties of the perturbative and non-perturbative subsectors in the large-$N$ limit \cite{Kurkela:2018dku}. 

To make this more precise, let us consider a dynamical system $\mathcal{F}$ defined on a fixed background metric  $ g^{(B)}_\mn $. The full system is composed of two subsectors, $\mathcal{U}$ and $\tilde{\mathcal{U}}$, representing the non-perturbative and perturbative degrees of freedom, respectively. These subsectors live in their own effective metric backgrounds $ g_\mn$ and $ \tilde{g}_\mn $. The effective action that captures the marginal/relevant deformations of each subsector and their mutual interaction can be written as~\cite{Kurkela:2018dku, Mondkar:2023gld}
\begin{eqnarray}\label{Eq:semi_action}
S_{{\rm tot}}[\phi, \tilde{\phi},g_\mn,\tilde{g}_\mn, g^{(B)}_\mn] &=& {S}[\phi,g_\mn] 
+ \tilde{{S}}[\tilde{\phi},\tilde{g}_\mn] \nonumber\\ &+& \int d^{d} x ~ \Big[ \frac{1}{2 \gamma} ~ \sqrt{-g^{(B)}}~ tr[(g-g^{(B)})\cdot {g^{(B)}}^{-1} \cdot(\tilde{g} - g^{(B)})\cdot {g^{(B)}}^{-1}] \nonumber \\
&+& \frac{\gamma'}{2 \gamma }  ~ ~\sqrt{-g^{(B)}} ~ \frac{(tr[g\cdot {g^{(B)}}^{-1}] - d ) (tr[\tilde{g}\cdot {g^{(B)}}^{-1}]-d)}{ \gamma'd-\gamma}  \Big]
\end{eqnarray}
in $d$-dimensional spacetime. Here ${S} $ and $\tilde{{S}} $ denote the effective actions of the subsectors, $ \mathcal{U}$ and $\tilde{\mathcal{U}}$, while $\phi $ and $ \tilde{\phi}$ collectively represent all matter fields in the two subsectors. The dimensionful parameters $ \gamma$ and $ \gamma '$ are semi-holographic couplings of mass dimension $ - d$; their ratio defines the dimensionless coupling $ \mathscr{R} := -\gamma^\prime/\gamma$. The effective metrics $ g_\mn$ and $\tilde{g}_\mn$ play the role of auxiliary fields and encode the marginal deformations of the respective subsectors. As mentioned earlier, the deformation is implemented by promoting $g_{\mu\nu}$ and $\tilde g_{\mu\nu}$ to local algebraic functionals of operators in the complementary subsector. {The couplings $\gamma$ and $\gamma'$ are of order $Q^{-d}$ with $Q$ the cut-off scale of the semi-holographic description, which is the saturation scale in the context of heavy-ion collisions \cite{Iancu:2014ava}}. 

The algebraic effective metric coupling equations follow from varying the action of Eq.~\eqref{Eq:semi_action} with respect to the effective metrics. Explicitly, these are
\begin{align} \label{Eq:metriccoup}
g_\mn &= g^{(B)}_\mn + \frac{\sqrt{- \tilde{g}}}{\sqrt{- g^{(B)}}}
\Big[ \gamma ~ \tilde{t}^{\alpha \beta} g^{(B)}_{\alpha \mu} g^{(B)}_{\beta \nu}   + \gamma' ~ tr(\tilde{t}\cdot g^{(B)}) g^{(B)}_\mn  \Big],\nonumber \\
\tilde{g}_\mn &= g^{(B)}_\mn + \frac{\sqrt{- {g}}}{\sqrt{- g^{(B)}}}
\Big[ \gamma ~ {t}^{\alpha \beta} g^{(B)}_{\alpha \mu} g^{(B)}_{\beta \nu}   + \gamma' ~ tr({t}\cdot g^{(B)}) g^{(B)}_\mn  \Big],
\end{align}
where $ t^{\alpha \beta}$ and $\tilde{t}^{\alpha \beta}$ are the energy-momentum tensors of the respective subsectors
\begin{eqnarray}
t^{\alpha \beta} = -\frac{2}{\sqrt{-g}} \frac{\delta {S}}{\delta g_{\alpha \beta}}, \hspace{1cm} \tilde{t}^{\alpha \beta} = -\frac{2}{\sqrt{-\tilde{g}}} \frac{\delta \tilde{{S}}}{\delta\tilde{g}_{\alpha \beta}}.
\end{eqnarray}

In the large-$N$ limit, expectation values of multi-trace gauge-invariant operators factorize. The metric-coupling equations \cref{Eq:metriccoup} can be interpreted as determining non-fluctuating effective metrics $g_{\mu\nu}$ and $\tilde g_{\mu\nu}$ in terms of the expectation values of the subsector energy-momentum tensors in any given state in the large-$N$ limit. Note that the effective metrics and the expectation values of the energy-momentum tensors of the subsystems should be determined together self-consistently. We will see how to do this explicitly later in this section.

Since the subsystem actions $S$ and $\tilde{S}$ are diffeomorphism invariant, their energy-momentum tensors are covariantly conserved with respect to their effective metric backgrounds, i.e., they satisfy the Ward identities 
\begin{align} \label{Eq:WI}
\nabla_\alpha t^{\alpha \beta} = 0, \hspace{1cm} \tilde{\nabla}_\alpha \tilde{t}^{\alpha \beta} = 0,
\end{align}
where $ \nabla$ and $ \tilde{\nabla}$ are the covariant derivatives with respect to $g_{\mu\nu}$ and $\tilde g_{\mu\nu}$, respectively.

The energy-momentum tensor of the full system $\mathcal{F}$ can be obtained as a polynomial of the energy-momentum tensors of the individual sectors by varying the action (Eq.~\eqref{Eq:semi_action}) with respect to the background metric $ g^{(B)}_\mn$
\begin{eqnarray} \label{Eq:fullem}
\mathcal{T}^{\mn} &:=& -\frac{2}{\sqrt{-g^{(B)}}} \frac{\delta {S}_{tot}}{\delta g^{(B)}_\mn},\nonumber \\
\mathcal{T}^\mu_\nu &=& \frac{1}{2} \Bigg( (t^\mu_{\phantom{\mu}\nu} + t^{\phantom{\nu}\mu}_\nu) \frac{\sqrt{-g}}{\sqrt{-g^{(B)}}} + (\tilde{t}^{\mu}_{\phantom{\mu}\nu} + \tilde{t}^{\phantom{\nu}\mu}_\nu) \frac{\sqrt{-\tilde{g}}}{\sqrt{-g^{(B)}}} \Bigg) + \Delta K \delta^\mu_\nu
\end{eqnarray}
where
\begin{eqnarray}\label{Eq:fullem1}
\Delta K = &-& \frac{\gamma}{2} ~ \Bigg( t^{\rho \alpha} \frac{\sqrt{-g}}{\sqrt{-g^{(B)}}} \Bigg) ~ g_{\alpha \beta}^{(B)} ~ \Bigg( \tilde{t}^{\beta \sigma} \frac{\sqrt{-\tilde{g}}}{\sqrt{-g^{(B)}}} \Bigg) ~ g^{(B)}_{\sigma \rho} \nonumber \\
&-& \frac{\gamma'}{2} ~ \Bigg( t^{\alpha \beta} \frac{\sqrt{-g}}{\sqrt{-g^{(B)}}} \Bigg) ~ g_{\alpha \beta}^{(B)} ~ \Bigg( \tilde{t}^{\sigma \rho} \frac{\sqrt{-\tilde{g}}}{\sqrt{-g^{(B)}}} \Bigg) ~ g^{(B)}_{\sigma \rho}.
\end{eqnarray}
The full energy-momentum tensor is explicitly a local algebraic functional of the subsystem energy-momentum tensors. One can verify explicitly that, provided that the coupling equations \cref{Eq:metriccoup} and the subsector Ward identities \cref{Eq:WI} are satisfied, the full energy-momentum tensor $ \mathcal{T}^\mu_\nu$ is covariantly conserved in the physical background $ g^{(B)}_\mn$  i.e., \cref{Eq:metriccoup} and \cref{Eq:WI} imply that
\begin{eqnarray} \label{Eq:TEM}
\nabla_{\mu}^{(B)} \mathcal{T}^\mu_\nu = 0.
\end{eqnarray}
As the full action $S_{\text{tot}}$ is not invariant under individual diffeomorphisms of the subsystems but only under their diagonal combination, the Ward identities of the individual subsectors should imply the Ward identity for the full system. Indeed, one can verify that \eqref{Eq:TEM} follows from \eqref{Eq:WI} by virtue of \eqref{Eq:metriccoup}. For practical purposes, $ g^{(B)}_\mn =\eta_{\mu\nu}$.

\subsection{The hybrid model and its thermal equilibrium}\label{sec:setup::subsec:model}

We specialize the general semi-holographic construction of \cref{sec:setup::subsec:action-formalism} to the concrete hybrid system used in this work. The full system consists of a weakly self-interacting conformal M\"uller–Israel–Stewart (MIS) fluid (perturbative/hard sector) coupled to a strongly self-interacting large-$N$ CFT with a holographic dual (non-perturbative/soft sector). Throughout, we work with $(2+1)$ boundary spacetime dimensions. Note that a finite-temperature $(2+1)$-dimensional large$-N$ CFT is holographically dual to a $(3+1)$ dimensional planar asymptotically Anti-de Sitter (aAdS$_4$) black brane. 

Our choice of a $(2+1)$-dimensional boundary theory is motivated both by phenomenology and by technical convenience. Many condensed-matter systems of current interest are effectively two-dimensional and admit effective descriptions in terms of  $(2+1)$-dimensional field theories, often involving an interplay between weakly and strongly interacting degrees of freedom.
Examples include Dirac materials such as graphene and topological-insulator surfaces, which host emergent massless Dirac fermions in two spatial dimensions~\cite{Novoselov:2005kj}. In many physically relevant regimes, these fermions interact with strongly fluctuating, and in some cases critical, collective modes~\cite{Hartnoll:2009sz, Giuliani_2012, Kosinski:2012vt}.
Another example of such systems is provided by $U(1)$ Dirac spin liquids in frustrated quantum magnets, whose low-energy physics is commonly captured by QED$_3$ with relativistic spinons coupled to an emergent gauge field~\cite{Nogueira:2005aj, Song:2018ccm, PhysRevX.14.021010, pjs3-14cc}. Likewise, many non-Fermi liquids such as strange metals are commonly modeled
by conducting band electrons coupled to degrees of freedom associated with
a quantum critical sector~\cite{PhysRevB.78.035103, Hartnoll:2009sz, PhysRevB.91.115111, PhysRevX.7.031058, Doucot:2024hzq}. Our semi-holographic construction, with a weakly interacting MIS sector coupled to a strongly interacting holographic CFT$_3$, is intended as a phenomenological
toy model for this broad class of hybrid $(2+1)$-dimensional systems. Working in aAdS$_4$ also keeps the gravitational problem technically simpler than in aAdS$_5$. In particular, this avoids the conformal anomaly present in the holographic sector, which complicates the inter-system coupling.\footnote{This complication can, however, be addressed via an iterative method as shown in \cite{Mukhopadhyay:2015smb}).}

Having specified the physical motivation and the structure of the two subsectors, we now describe the thermal equilibrium configuration of the semi-holographic model. In the next subsection, we will describe the perturbations 
that define its hybrid collective modes. Our focus throughout is on linear response about global thermal equilibrium. In thermal equilibrium, the holographic (soft) sector is dual to a stationary AdS$_4$–Schwarzschild black brane. In ingoing Eddington–Finkelstein coordinates the metric, $G_{MN}$, of stationary AdS$_4$–Schwarzschild black brane takes the form
\begin{align}
    ds^2 =& - a^2 A(r) dt^2 + 2 a \:dr dt + \frac{b^2 r^2}{L^2} \left(dx^2 + dy^2 \right)\label{eqn:bulk-metric-static-AdS4},\\
    A(r) &= \frac{r^2}{L^2} - \frac{\Mbh L^2}{r},
\end{align}
where $\Mbh$ is the Arnowitt–Deser–Misner (ADM) mass of black brane~\cite{PhysRev.116.1322, Arnowitt:1962hi, Lu:1993vt} which is the solution of pure Einstein's gravity in four dimensions with a negative cosmological constant.
Note that $(t,\vec{x}) = (t,x,y)$ are boundary spacetime coordinates, whereas $r$ is the bulk radial coordinate. The black hole horizon is at radial location $r = r_h = L^{4/3} \Mbh^{1/3}$.  
The spacetime boundary is at $r \to \infty$. $L$ is the AdS$_4$ radius, which we henceforth set to unity. As $r \rightarrow \infty$, we see that the boundary metric is given by 
\begin{equation}\label{AdS4-eff-boundary}
   g_{\mn} = \text{diag} \left(-a^2, b^2, b^2  \right),
\end{equation}
which is the effective metric background in which the soft-sector energy-momentum tensor, $t^\mn$, is defined. At equilibrium, $a$ and $b$ are constants. The energy-momentum tensor of the soft sector can then be obtained using holographic renormalization~\cite{Balasubramanian:1999re, deHaro:2000vlm, Skenderis:2002wp} {and explicitly is}
\begin{align}\label{eq:MIS-tmunu-eqlm}
t^\mn =(\epsilon+p)u^\mu u^\nu  +p g^\mn ,
\end{align}
{with $g_{\mn}$ being the effective metric \eqref{AdS4-eff-boundary}},
\begin{equation}
    u^\mu = \left(\frac{1}{a}, 0, 0 \right),
\end{equation}
and $\epsilon = 2p = 2 n_1 N^2 T_1^3$ with $n_1 = \frac{32 \pi^3}{27}$ and $T_1 = \frac{3}{4 \pi}  \Mbh^{1/3}$ {(by the holographic dictionary $N^2 =\frac{1}{8 \pi G_N^{(4)}}$.)}

The conformal MIS fluid forms the perturbative (hard) subsector. Its energy-momentum tensor at equilibrium takes the perfect fluid form,
\begin{align}\label{eq:MIS-tmunu-eqlm}
\tilde{t}^\mn =(\et+\pt)\Ut^\mu \Ut^\nu  +\pt \gt^\mn ,
\end{align}
which is conserved, $\tilde\nabla_\mu \tilde{t}^\mn=0,$ with respect to its own effective metric,
\begin{equation}\label{MIS-eff-metric}
    \gt_{\mn} =  \text{diag} \left(-\At^2, \Bt^2, \Bt^2  \right).
\end{equation}
At equilibrium $\At$ and $\Bt$ are constants. The equilibrium fluid four-velocity $\Ut^\mu$ is given by
\begin{equation}
    \Ut^\mu = \left(\frac{1}{\At}, 0, 0 \right)
\end{equation}
which is timelike normalized, $\Ut^\mu \Ut_\mu = -1$. In (2+1)-dimensions, the conformal equation of state (EoS) is $\et = 2 \pt = 2 n_2 {N^2} T_2^3 $, where $\et$ and $\pt$ denote the energy density and pressure, respectively, and $T_2$  is the temperature of the MIS sector.

The complete system resides in the physical background metric $g^{(B)}_{\mu \nu}$, which we take to be flat Minkowski spacetime
\begin{equation}
    g^{(B)}_{\mu \nu} = {\rm diag} \left(-1,1,1 \right)
\end{equation}
The two sectors are coupled with each other via the algebraic effective metric coupling equations given in  \cref{Eq:metriccoup} which self-consistently determine $a$, $b$, $\At$, $\Bt$ given the expressions of equilibrium energy-momentum tensors of each sector. {Explicitly, \cref{Eq:metriccoup} is of the form}
\begin{subequations}
\begin{align}
    1 -a^2 &=  \gamma \left(  \frac{ \Bt^2 }{\At} \et + \mathscr{R} \frac{\left(\At^2 -\Bt^2 \right) }{\At} \et \right), \\
    1 - b^2 &= \gamma \left( \frac{\At}{2} \et - \mathscr{R} \frac{ \left( \At^2 -\Bt^2  \right) }{\At}  \et \right), \\
    1 - \At^2 &= \gamma \left( \frac{b^2}{a} \varepsilon +  \mathscr{R} \frac{ \left( \A^2 -\B^2  \right) }{\A} \varepsilon \right),   \\
    1 - \Bt^2 &= \gamma \left( \frac{\A}{2} \varepsilon -  \mathscr{R} \frac{ \left( \A^2 -\B^2  \right)}{ \A}  \varepsilon \right)
\end{align}
\end{subequations}
{To find the solution corresponding to thermal equilibrium of the full semi-holographic system characterized by the physical temperature $T$, the subsystem temperatures $T_1$ and $T_2$ should be determined self-consistently via the global equilibrium condition} \cite{Kurkela:2018dku, Mitra:2025afp,  Mondkar:2023gld}
\begin{align}\label{eq:thermal-eq}
    T = a T_1 = \At T_2.
\end{align}
{We can find physical solutions of the full system corresponding to thermal equilibrium at arbitrary values of the physical temperature $T$ when $\gamma >0$ and $\mathscr{R} >1$ \cite{Kurkela:2018dku}.}

As mentioned earlier in the Introduction, in semi-holography, there exist so-called pseudo-equilibrium states in the large-$N$ limit, characterized by two equilibrated subsectors at respective physical temperatures $T_1$ and $T_2$ that do not satisfy the global equilibrium condition \cref{eq:thermal-eq}. However, as discussed in the Introduction, it has been proven in~\cite{Mitra:2025afp} that the global thermal equilibrium, characterized by  \cref{eq:thermal-eq} maximizes the total entropy in the micro-canonical ensemble (which is not obvious in the effective metric coupling scheme) and the full isolated system relaxes to the global thermal equilibrium state in the limit when the average energy density is infinite (i.e. large in the scale of the inter-system coupling) for typical initial conditions even in the large-$N$ limit.

\section{Computation of the semi-holographic quasinormal modes}\label{sec:methodology}

Quasinormal modes (QNM) are collective modes of dissipative systems that govern the linear response near thermal equilibrium. In holographic systems, the QNM at a finite temperature are described by the linearized normalizable fluctuations of an AdS-Schwarzchild black brane with ingoing boundary conditions at the horizon \cite{ Starinets:2002br, Kovtun:2005ev, Ammon:2015wua}. In semi-holography, the computation of QNM is more complicated as the boundary value of the bulk field, which is identified with the source of the dual operator, is dynamical and should be solved self-consistently together with the fluctuations of the perturbative sector living at the boundary. This has been implemented earlier in the context of a bulk scalar interacting with a boundary Goldstone mode in \cite{Mondkar:2021qsf}. In this work, we extend the method developed in \cite{Mondkar:2021qsf} to the case of the semi-holographic setup with effective metric couplings described in the previous section. This is technically much more involved.

Here, we consider the linearized perturbations of both the AdS$_4$-Schwarzchild bulk metric \eqref{eqn:bulk-metric-static-AdS4} describing the thermal holographic sector and the variables of the hard MIS subsector, which are coupled via their effective metrics. The non-equilibrium perturbations of the bulk metric give rise to perturbations of both the energy-momentum tensor of the soft sector and the boundary effective metric \eqref{AdS4-eff-boundary}, which should be determined self-consistently together with those of the hard (MIS) subsector. The perturbed effective merics of the subsectors will be dynamical, inhomogeneous, and also have off-diagonal components. We discuss the fluctuations in the shear and sound channels separately (these are the only two channels in the $2+1$-dimensional field theory).\footnote{The graviton in AdS$_4$ has two degrees of freedom. The {equilibrium} black brane breaks the full conformal invariance but retains the rotational (SO(2)) and translational invariance. The shear and sound channels describe modes that transform as vectors and scalars under the residual O(1) at any finite momentum.}  

Before describing linearized fluctuations about thermal equilibrium, it is pertinent to describe the dynamics of the full system explicitly. The coarse-grained dynamics of the perturbative sector is given by the MIS equations, which are as follows. The energy-momentum tensor of the perturbative sector is first parameterized in the form
\begin{align}\label{emt-mis}
\tilde{t}^\mn =(\et+\pt)\Ut^\mu \Ut^\nu  +\pt \gt^\mn + \tilde{\Pi}^\mn,
\end{align}
where {recall from previous section that} $\et = 2 \pt$ is given in terms of the temperature $T_2$ by the EoS $\et = 2 n_2 N^2 T_2^2$, and $\tilde{\Pi}^\mn$ is the dissipative part of the energy-momentum tensor satisfying
\begin{equation}
    \tilde{\Pi}^\mn \Ut_\mu = 0
\end{equation}
so that $\Ut^\mu$ is identified with the velocity of energy flow (the so-called Landau-Lifshitz frame). The dynamical equations for $T_2$, $\Ut^\mu$ and $\tilde{\Pi}^\mn$ are given by 
\begin{subequations}\label{Eq:MIS}
\begin{align}
    &\tilde{\nabla}_\alpha \tilde{t}^{\alpha \beta} = 0, \label{Eq:MIS-sub1}\\
    &\left( \taut_\pi \Ut^{\alpha} \nabt_{\alpha} + 1 \right) \Pit^{\mn} = -\etat \sigt^{\mn}, \label{Eq:MIS-sub2}
\end{align}
\end{subequations}
where $\taut_\pi$ is the relaxation time and $\etat$ is the shear viscosity. We will use the following dimensionless parameterizations of the transport coefficients: $T_2 \etat = \Cet (\et + \pt)$, and $ T_2\taut_\pi =\Ctaut$. We will choose $\Cet = 10/4 \pi$ and $\Ctaut = 5 \Cet$. This choice of transport coefficients is motivated by the fact that the MIS theory models a \textit{weakly interacting} hard sector. Note that the relation $\Ctaut = 5 \Cet $ is found in the relaxation time approximation of kinetic theory, see e.g.~\cite{Romatschke:2017ejr}. Note that the dynamical equations \eqref{Eq:MIS} are in the effective background $\gt_\mn$ which should be determined self-consistently as described below.

The bulk metric fluctuations should satisfy Einstein's equations
\begin{equation}\label{eq:bulk-einstein}
    R_{MN} - \frac{1}{2} R G_{MN} - 3 G_{MN} =0.
\end{equation}
with a negative cosmological constant in four spacetime dimensions (recall we have set $L$, the radius of AdS, to unity). In semi-holography, the boundary metric $g_\mn$ (given by $g_\mn = \lim_{r\rightarrow\infty} r^{-2}G_{\mu\nu}$), which is identified with the effective metric of the holographic sector, should be determined self-consistently. This is in contrast to holography, where the boundary metric is identified with the physical non-dynamical background metric $g^{(B)}_{\mu\nu}$.

Finally, both the effective metrics $g_\mn$ and $\gt_\mn$ are determined together via the metric coupling equations \eqref{Eq:metriccoup} for a given physical non-dynamical background metric $g^{(B)}_{\mu\nu}$ which we choose to be the Minkowski metric, $\eta_{\mu\nu} = \rm diag \left(-1,1,1 \right)$. 

Clearly, it is formidable to solve the full non-linear dynamics, which has, however, been achieved successfully using iterative methods when the coupling of the two sectors involves effective sources of scalar operators \cite{Mukhopadhyay:2015smb,Mondkar:2021qsf,Ecker:2018ucc}. In what follows, we will describe how to solve the linearized fluctuations of the full system about thermal equilibrium separately in the shear and sound channels. The reader uninterested in these details may skip the rest of this section and go to the results presented in the following section.

\subsection{Shear channel}
We denote the linearized perturbations of the equilibrium bulk metric~\eqref{eqn:bulk-metric-static-AdS4} by $\delta G_{MN}$, which admit the following Fourier decomposition in the boundary spacetime domain owing to the translational invariance in the boundary field theory,
\begin{equation}\label{eq:bulk-metric-pert-Fourier}
    \delta G_{MN}(r, t, \vec{x}) = \int d^2 \vec{q} \int d \omega e^{\mathrm{i} (-\omega t + \vec{q}.\vec{x})}  \delta G_{MN} (r, \omega, \vec{q})
\end{equation}
here $\omega$ and $\vec{q}$ denote the Fourier frequency and momentum, respectively. By exploiting the $SO(2)$ rotational invariance in the boundary spatial domain, we can align the plane-wave perturbation along the y-axis direction, i.e., $\vec{q} = (0, q)$. 
In what follows, we will suppress the dependence on $(\omega, q)$ from the argument for compactness.

The individual perturbations $\delta G_{MN}$ are not invariant under bulk-diffeomorphisms and thus not physical. However, one can construct a diffeomorphism-invariant bulk perturbation as a linear combination of perturbations $\delta  G_{MN}$. We denote the diffeomorphism-invariant perturbation in the shear channel by $\Zsh(r)$, which is given by
\begin{equation}\label{eq:Zsh}
    Z_{sh}(r) := \omega \delta G_{yx}(r) + q \delta G_{t x}(r)
\end{equation}
$Z_{sh}(r)$ obeys a second-order ODE, 
\begin{small}
\begin{align}\label{eq:gauge-inv-var-shear-holo-bulk}
    0&= \frac{ b^2  r^2  \left(27 a^3 r^3 - 64 \pi^3 T^3 \right) \Big( \left(64 \pi^3 T^3-27 a^3  r^3 \right) \momentum^2 + 27 a b^2 r^3 \freq^2 \Big) }{729 a^3}Z_{sh}''(r)  \nonumber \\ 
    & + \frac{2 \mathrm{i} b^2 \pi r^3 T \freq \left( a \left( 27 a^3 r^3 - 64 \pi^3 T^3 \right)\momentum^2 - 96 \mathrm{i} b^2 \pi^2 r T^2 \freq  -27 a^2 b^2 r^3 \freq^2\right) }{27 a^2} Z_{sh}'(r)   \nonumber \\
    & +   \Big( - a^3 \momentum^2 \left(\frac{64 \pi^3 T^3}{27 a^3 } - r^3 \right) \left(\pi^2 \momentum^2 r T^2 + 2 b^2 \left( r^3 - \frac{64 \pi^3 T^3}{27 a^3 } \right) \right) - \mathrm{i} \pi a^2 b^2 \momentum^2 r^2 T \left( \frac{64 \pi^3 T^3}{27 a^3 }  + 2 r^3 \right) \freq \nonumber \\ 
    & - a b^2 r^3 \left( \pi^2 \momentum^2 r T^2 + 2 b^2 \left( \frac{128 \pi^3 T^3}{27 a^3 } + r^3 \right) \right) \freq^2 + 2 \mathrm{i} \pi b^4 r^5 T \freq^3  \Big)  Z_{sh}(r),
\end{align}
\end{small}
which follows from the linearized bulk Einstein equations. $'$ denotes derivative with respect to the argument. Here we have introduced the dimensionless frequency, $\freq  = \omega/(\pi T)$ and dimensionless momentum, $\momentum = q /(\pi T)$. Recall that $T$ is the global thermal equilibrium temperature of the full semi-holographic system.

In the MIS sector, the perturbations appear in $\Ut^\mu$ and $\tilde{\Pi}^\mn$. In particular, the perturbed $\Ut^\mu$ is
\begin{equation}
    \Ut^\mu = \left(\frac{1}{\At}, \vt,0 \right)
\end{equation}
whereas one independent perturbation $\pityx$ shows up in $\tilde{\Pi}^\mn$. At the same time, perturbations $\delta \gttx$ and $\delta \gtyx$ are turned on in the effective metric $\gt_\mn$ of the MIS sector. We emphasize that after perturbing the system, the diagonal elements of both $g_\mn$ and $\gt_\mn$ become spacetime dependent. We denote the perturbations in the effective metrics $g_\mn$ and $\gt_\mn$ by $\delta g_\mn$ and $\delta \gt_\mn$, respectively. After linearizing the MIS equations of motion~\eqref{Eq:MIS} in perturbations $\vt$, $\pityx$, $\delta \gttx$ and $\delta \gtyx$ we get the following two linearized equations,
\begin{align}\label{eq:MIS-shear-hydro-eom-lin}
    \Bt^2 \At^5 \momentum \pityx  - 3 \nt T^3 \left( \At \Bt^2 \vt + \delta \gttx \right) \freq = 0
\end{align}
\begin{equation}\label{eq:MIS-shear-Pi-eom-lin}
    \At^3 \Bt^4 \left(\mathrm{i}  + 5  \Cet \pi \freq\right) \pityx  - 3  \nt \pi T^3 \Cet \left( \At \Bt^2 \momentum \vt - \freq \delta \gtyx \right) = 0
\end{equation}
The first one follows from the conservation equation for $\tilde{t}^\mn$, Eq.~\eqref{Eq:MIS-sub1}, whereas the second one follows from the relaxation equation for $\tilde{\Pi}^\mn$, Eq.~\eqref{Eq:MIS-sub2}.

The effective metric perturbations $\delta \gtx$, $\delta \gyx$, $\delta \gttx$, $\delta \gtyx$, which are active in the shear channel, satisfy linearized versions of effective metric coupling equations~\eqref{Eq:metriccoup}. These linearized metric coupling equations are explicitly as follows,
\begin{small}
\begin{subequations}\label{eq:shear-coup-lin}
\begin{align}
    \D \gtx &= -\gamma \frac{\left( 3 \At \Bt^2  \vt  + \D \gttx \right)\nt T^3}{\At^4}  \\
    \D \gyx &= \gamma \left( - \frac{\nt T^3 }{\At^2 \Bt^2} \D \gtyx + \pityx  \right)   \\
   \left(-b^2 \freq^2 + a^2 \momentum^2 \right) \D \gttx &=  \gamma \Big(  \frac{64 \pi^3 \momentum^2 T^3 \D \gtx }{ 27 a^2} - \frac{3 \mathrm{i} \pi^3 \momentum^4  T^3 \D \gtx}{4 b^2} \freq  - \frac{\pi^3 T^3 \left( 256 b^4 \D \gtx + 81 \mathrm{i} a^4 \momentum^3 \D \gyx \right)}{108 a^4 b^2} \freq^2 \nonumber \\
   &+\frac{ \mathrm{i} \pi^3 \momentum^2 T^3 \D \gtx }{2 a^2} \freq^3 + \frac{\mathrm{i} \pi^3 \momentum T^3 \D \gyx}{2 a^2 } \freq^4 + \frac{3 a \momentum}{2 \pi T} \zshn \Big) \\
   \left(-b^2 \freq^2 + a^2 \momentum^2 \right) \D \gtyx &= \gamma \Big(  -\frac{32 \pi^3 \momentum^2 T^3 \D \gyx }{ 27 b^2}  + \frac{\pi^3 T^3}{108 } \left( \frac{81 \mathrm{i}\momentum^3 \D \gtx}{b^2} + \frac{128 \D \gyx}{a^2}\right)\freq^2  \nonumber \\
   & + \frac{3 \imag \pi^3 \momentum^2 T^3 \D \gyx  }{4 b^2} \freq^3 - \frac{\imag \pi^3 \momentum T^3 \D \gtx}{2 a^2}\freq^4  - \frac{\imag \pi^3 T^3 \D \gyx }{2 a^2} \freq^5 - \frac{3 a \freq}{2 \pi T} \zshn \Big)
\end{align}
\end{subequations}
\end{small}
where $\zshn$ is the normalizable mode of $\Zsh(r)$ which can be expressed in terms of normalizable modes of $\delta G_{tx}$ and $\delta G_{yx}$ as shown in Eq.~\eqref{eq:zsh3} of App.~\ref{app:near-boundary-exp::subsec-shear}.

The hybrid QNM frequencies in the shear channel are values of $\freq(\momentum)$ for which a non-trivial solution exists to the set of seven coupled equations~\eqref{eq:gauge-inv-var-shear-holo-bulk}, \eqref{eq:MIS-shear-hydro-eom-lin}, \eqref{eq:MIS-shear-Pi-eom-lin}, and \eqref{eq:shear-coup-lin} in seven variables $\Zsh(r)$, $\vt$, $\pityx$, $\delta \gtx$, $\delta \gyx$, $\delta \gttx$, and $\delta \gtyx$. The hybrid QNM spectrum of the semi-holographic model can be obtained numerically by adapting the standard pseudospectral procedure used in holography~\cite{Yaffe-2014, Jansen:2017oag}. In conventional holographic setups, the problem reduces to a linear eigenvalue equation. However, in semi-holography once the dynamical boundary conditions are introduced, the spectrum instead follows from a polynomial eigenvalue equation, which smoothly approaches the familiar linear form as the sectors decouple ($\gamma \to 0$).  We also refer the reader to Appendix A of our previous work~\cite{Mondkar:2021qsf}, which details the methodology of the numerical QNM computation in a simpler semi-holographic model employing scalar couplings rather than effective metric couplings.

We define a dimensionless radial coordinate $u = r_h/r$ which has a finite range such that the spacetime
boundary is at $u=0$ and the horizon is at $u=1$. {Recall $r_h = \Mbh^{1/3} = 4 \pi / (3 T_1) =  4 \pi a / (3 T) $ from previous section.}
The bulk radial domain of interest, $r \in [0, \infty)$, gets mapped to a finite domain $u \in [0,1]$. From the structure of the near boundary expansion,  \cref{eq:Zsh-near-boundary-exp}, it is clear that we can define a new function $g_{sh}(u, q, \omega)$ via 
\begin{equation}
g_{sh}(u)   := \frac{r_h}{u} \Big( \Zsh(u)  -\Big( \left(\omega \D \gyx + q \D \gtx \right)\frac{r_h^2}{u^2}  - \frac{\mathrm{i} \omega}{a} \left(\omega \D \gyx + q \D \gtx \right) \frac{r_h}{u} - \frac{q^2}{2 b^2} \left(\omega \D \gyx + q \D \gtx \right)  \Big) \Big)
\end{equation}
which is regular at $u=0$. For brevity, we suppress the $q, \omega$ labels. This definition guarantees that $g_{sh}(u)$ is analytic at $u=1$, consistent with the ingoing boundary condition at the black hole horizon. Furthermore for $u = 0$, $g_{sh}(u)$ becomes exactly the normalizable mode of $\Zsh(u)$.
The above definition of $g_{sh}(u)$ ensures a smooth decoupling limit as $\gamma \to 0$. 
We
can readily obtain the differential equation for $g_{sh}(u)$ by substituting the above in \cref{eq:gauge-inv-var-shear-holo-bulk}.

To solve numerically, we expand $g_{sh}(u)$ in the first $K+1$ Chebyshev polynomials, which are linearly mapped onto the interval $[0,1]$. The $K+1$ collocation points for the pseudospectral grid are obtained from the Gauss–Lobatto nodes $\cos(n\pi/K)$ of the interval $[-1,1]$, linearly mapped to the interval $[0,1]$ as $(1-\cos(n\pi/K))/2$ for $n=0,\dots,K$~\cite{Boyd-Chebyshev}. Substituting the truncated series into the differential equation for $g_{sh}(u)$ and evaluating at each collocation point yields $K+1$ algebraic relations for the $K+1$ expansion coefficients.

In addition to these $K+1$ equations, which descend from the bulk equation~\eqref{eq:gauge-inv-var-shear-holo-bulk}, the rest of the six equations, namely the two MIS equations~\eqref{eq:MIS-shear-hydro-eom-lin}, \eqref{eq:MIS-shear-Pi-eom-lin} and four metric coupling equations~\eqref{eq:shear-coup-lin} for the six variables $\vt$, $\pit_{yx}$, $\D \gtx$, $\D \gyx$, $\D \gttx$, $\D \gtyx$ implement the dynamical boundary conditions. Together, this produces a coupled system of $K+7$ equations which can be organized as a single
matrix equation of the form 
\begin{equation}\label{eq:fifth-order-eigenvalue-problem-shear}
    \hat{Q}\cdot V = 0
\end{equation}
where $K+7$ dimensional column vector $V$ contains all the $K+7$ variables. $\hat{Q}$ is a $K+7$ dimensional square matrix. A non-trivial solution exists to  \cref{eq:fifth-order-eigenvalue-problem-shear} only for certain values of $\freq$ at which $\det \hat{Q} = 0$.
These values of $\freq$ constitute the QNM spectrum of the full system. However, one can follow a
more efficient strategy than solving for $\det \hat{Q} = 0$ to determine the QNM
spectrum.

We note that each element of $\hat{Q}$ involves at most the fifth power of $\freq$. Therefore, we can write $\hat{Q} = Q_0 + \freq Q_1 + \freq^2 Q_2 + \freq^3 Q_3 + \freq^4 Q_4 + \freq^5 Q_5$ where $Q_0$, $Q_1$, $Q_2$, $Q_3$, $Q_4$, $Q_5$ are independent of $\freq$. Thus  \cref{eq:fifth-order-eigenvalue-problem-shear} is simply a polynomial eigenvalue problem of fifth order whose solutions give us the desired QNM frequencies for a given
value of $\momentum$.

We can readily solve a fifth-order polynomial eigenvalue problem by converting it into a higher-dimensional generalized
linear eigenvalue problem. To see this, we note that we can rewrite  \cref{eq:fifth-order-eigenvalue-problem-shear} in the form

\begin{subequations}\label{eq:generalized-eigenvalue-problem-shear}
\begin{equation}\label{eq:generalized-eigenvalue-problem-shear-sub1}
    P \begin{pmatrix}
V  \\
\freq V  \\
\freq^2 V \\
\freq^3 V \\
\freq^4 V 
\end{pmatrix} = \freq R \begin{pmatrix}
V  \\
\freq V  \\
\freq^2 V \\
\freq^3 V \\
\freq^4 V 
\end{pmatrix} 
\end{equation}

\begin{equation}
P = \begin{pmatrix}
O & I & 0 & 0 & 0 \\
0 & 0 & I & 0 & 0 \\
0  & 0 & 0 & I & 0 \\
0  & 0 & 0 & 0 & I \\
Q_0 & Q_1 & Q_2 & Q_3 & Q_4
\end{pmatrix},  
\qquad 
R = \begin{pmatrix}
I & 0 & 0 & 0 & 0 \\
0 & I & 0 & 0 & 0 \\
0  & 0 & I & 0 & 0 \\
0  & 0 & 0 & I & 0 \\
0 & 0 & 0 & 0 & - Q_5
\end{pmatrix}  
\end{equation}
\end{subequations}
where $O$ is the $(K+7)$ dimensional null matrix and $I$ is the identity matrix with the same rank as $\hat{Q}$.
Note that  \cref{eq:fifth-order-eigenvalue-problem-shear} is $(K+7)$ dimensional fifth-order polynomial eigenvalue problem, whereas,  \cref{eq:generalized-eigenvalue-problem-shear-sub1} is a generalized linear eigenvalue problem in $5(K+7)$ dimensions. We
can obtain the full QNM spectrum by using standard routines for solving the generalized
eigenvalue problem of the form  \cref{eq:generalized-eigenvalue-problem-shear-sub1}.

\subsection{Sound channel}

In the sound channel, the bulk diffeomorphism-invariant perturbation is
\begin{align}\label{eq:Zso}
    Z_{so}(r) &:= q^2 \delta G_{tt}(r) + \omega^2 \delta G_{yy} + 2 q \omega \delta G_{t y}   \nonumber \\
    &+ q^2 \left( \frac{a^2}{b^2} + \frac{a^2}{b^2} \frac{\Mbh}{2 r^3}  - \frac{\omega^2}{q^2}\right) \delta G_{xx}(r)
\end{align}
$\Zso(r)$ satisfies a linearized second-order ODE, which follows from the bulk Einstein equations,
\begin{small}
\begin{align}\label{eq:gauge-inv-var-sound-holo-bulk}
    0&=\frac{16 b^2 r^2 \left(  27 a^3 r^3 - 64 \pi^3 T^3 \right) \Big(  \left( 27 a^3 r^3 - 16 \pi^3 T^3 \right)\momentum^2 -27 a b^2 r^3 \freq^2  \Big)}{729 a^2} Z_{so}''(r)   \nonumber \\
    & + \frac{32 \pi  b^2 r T }{243 a^2} \Big( 16 \pi  T^2 \left(32 \pi^3 T3 + 27 a^3 r^3 \right) \momentum^2 - 9 \mathrm{i} a^2 r^2 \left( 27 a^3 r^3 - 16 \pi^3 T^3 \right) \momentum^2 \freq \nonumber \\
    &-864 \pi^2  a b^2 r^3  T^2 \freq^2 + 243 \mathrm{i} a^3 b^2 r^5 \freq^3 \Big) Z_{so}'(r)   \nonumber \\
    & + 4 a \momentum^2  \Big( a^3  \left( \frac{64 \pi^3 T^3 }{27 a^3} - 4 r^3  \right) \left(\pi^2  r T^2 \momentum^2 +  b^2 (\frac{64 \pi^3 T^3 }{27 a^3} + 2 r^3) \right) + 4 \pi \mathrm{i} a^2 b^2  r^2 T   \left( \frac{64 \pi^3 T^3 }{27 a^3} + 2 r^3 \right) \momentum \freq  \nonumber \\
    & + 4 a b^2 r^3 \left( \pi^2 r T^2 \momentum^2  +  2 b^2 \left( \frac{128 \pi^3 T^3 }{27 a^3} + r^3  \right) \right) \freq^2 - 8 \pi \mathrm{i} b^4 r^5 T \freq^3 \Big)  Z_{so}(r)
\end{align}
\end{small}
In the sound channel of the MIS sector, there are three independent perturbations $\vt$, $\pitxx$, and $\delta T_2$ which follow three coupled linearized equations of motion. Two of those equations come from the conservation equation~\eqref{Eq:MIS-sub1}, which are,
\begin{subequations}\label{eq:MIS-sound-hydro-lin}
\begin{align}
    2 \At  \Bt^2 \left( \momentum T \vt - 2 \freq \delta T_2   \right) - \freq T \left( \delta \gtxx + \delta \gtyy \right) &= 0  \\
    2 \At^5 \Bt^2 \pi \freq T \pitxx + 3 \nt \pi T^3 \left( 2 T \freq \left( \At \Bt^2 \vt + \delta \gtty \right) + \momentum \left( T \delta \gttt - 2 \At^3 \delta T_2 \right)  \right) &=0 
\end{align}
\end{subequations}
whereas the third equation follows from the relaxation equation~\eqref{Eq:MIS-sub2}, which is,
\begin{align}\label{eq:MIS-sound-Pi-eqn-lin}
    2 \At^3 \Bt^4 \left( \mathrm{i}  + 5 \Cet \pi \freq \right) \pitxx + 3 \nt \pi T^3 \Cet \left( 2 \At \Bt^2 \momentum \vt +  \left( \delta \gttx - \delta \gtty \right)\freq \right)  = 0
\end{align}
There are eight active effective metric perturbations $\D \gtt$, $\D \gxx$, $\D \gyy$, $\D \gty$, $\D \gttt$, $\D \gtxx$, $\D \gtyy$, $\D \gtty$ in the sound channel which obey eight linearized metric coupling equations obtained from~\eqref{Eq:metriccoup} by linearizing in all the perturbations. In all, the complete set of equations of the sound channel constitute twelve coupled linearized equations, namely Eqs.~\eqref{eq:gauge-inv-var-sound-holo-bulk},~\eqref{eq:MIS-sound-hydro-lin},~\eqref{eq:MIS-sound-Pi-eqn-lin} and eight linearized metric coupling equations, in twelve variables $\Zso(r)$, $\vt$, $\pitxx$, $\delta T_2$, $\D \gtt$, $\D \gxx$, $\D \gyy$, $\D \gty$, $\D \gttt$, $\D \gtxx$, $\D \gtyy$, $\D \gtty$. The set of frequencies $\freq(\momentum)$ enabling non-trivial solutions to these twelve coupled equations constitutes the QNM spectrum of the full semi-holographic system in the sound channel.

The computation of the hybrid QNM in the sound channel follows analogously to the shear channel computation described in the previous subsection.
 As evident from the near boundary expansion of \cref{eq:Zso-near-boundary-exp}, in the case of the sound channel, we can define a function $g_{so}(u, q, \omega)$  
 via 
\begin{align}
    g_{so}(u) &:=   \frac{r_h}{u} \Big( \Zso(r) - \left( \omega^2 \left( \D \gyy - \D \gxx \right) + 2 \omega  q \D \gty + q^2 \left(\D \gtt +\frac{a^2 \D \gxx}{b^2} \right)  \right) \frac{r_h^2}{u^2} \nonumber \\
    &+  \frac{\imag \omega }{a}  \left( \omega^2 \left( \D \gyy - \D \gxx \right) + 2 \omega  q \D \gty + q^2 \left(\D \gtt +\frac{a^2 \D \gxx}{b^2} \right)  \right)  \frac{r_h}{u}  \nonumber \\
    & + \frac{q^2}{2 b^2}  \left( \omega^2 \left( \D \gyy - \D \gxx \right) + 2 \omega  q \D \gty + q^2 \left(\D \gtt +\frac{a^2 \D \gxx}{b^2} \right)  \right)   \Big).
\end{align}
which is regular at $u=0$ by definition. We can readily discretize the bulk equation for $g_{so}(u)$ on a Gauss-Lobatto grid by approximating it as a linear combination of the first $K+1$ Chebyshev polynomials. Evaluating it on $K+1$ grid points yields $K+1$ equations in terms of $K+1$ expansion coefficients. The dynamical boundary conditions are specified by eight algebraic effective metric coupling equations together with three equations~\eqref{eq:MIS-sound-hydro-lin}, and~\eqref{eq:MIS-sound-Pi-eqn-lin} of the MIS sector in terms of eleven variables $\D \gtt$, $\D \gxx$, $\D \gyy$, $\D \gty$, $\D \gttt$, $\D \gtxx$, $\D \gtyy$, $\D \gtty$, $\vt$, $\D T_2$, and $\pit_{xx}$. The resulting coupled system of $K + 12$ equations (in terms of $K+12$ variables) forms a seventh-order polynomial eigenvalue problem for $\freq$ of the form 
\begin{equation}\label{eq:fifth-order-eigenvalue-problem-sound}
    \hat{\mathcal{Q}}\cdot \mathcal{V} = 0
\end{equation}
We note that each element of $\hat{\mathcal{Q}}$ involves at most the seventh power of $\freq$. Therefore, we can write $\hat{\mathcal{Q}} = \mathcal{Q}_0 + \freq \mathcal{Q}_1 + \freq^2 \mathcal{Q}_2 + \freq^3 \mathcal{Q}_3 + \freq^4 \mathcal{Q}_4 + \freq^5 \mathcal{Q}_5 + \freq^6 \mathcal{Q}_6 + \freq^7 \mathcal{Q}_7 $ where $\mathcal{Q}_0$, $\mathcal{Q}_1$, $\mathcal{Q}_2$, $\mathcal{Q}_3$, $\mathcal{Q}_4$, $\mathcal{Q}_5$, $\mathcal{Q}_6$ , $\mathcal{Q}_7$  are independent of $\freq$. This seventh-order polynomial eigenvalue problem in $K+12$ dimensions can be converted to a generalized linear eigenvalue problem in $7 (K+12)$ dimensions of the form

\begin{align}\label{eq:generalized-eigenvalue-problem-sound}
    \mathcal{P}  \begin{pmatrix}
\mathcal{V}  \\
\freq \mathcal{V}  \\
\freq^2 \mathcal{V} \\
\freq^3 \mathcal{V} \\
\freq^4 \mathcal{V}  \\
\freq^5 \mathcal{V}  \\
\freq^6 \mathcal{V} 
\end{pmatrix} 
&= \freq \mathcal{R} \begin{pmatrix}
\mathcal{V}  \\
\freq \mathcal{V}  \\
\freq^2 \mathcal{V} \\
\freq^3 \mathcal{V} \\
\freq^4 \mathcal{V}  \\
\freq^5 \mathcal{V}  \\
\freq^6 \mathcal{V} 
\end{pmatrix} ,\\
\mathcal{P} = \begin{pmatrix}
O & I & 0 & 0 & 0 & 0 & 0 \\
0 & 0 & I & 0 & 0 & 0 & 0\\
0  & 0 & 0 & I & 0 & 0 & 0\\
0  & 0 & 0 & 0 & I & 0 & 0\\
0  & 0 & 0 & 0 & 0 & I & 0\\
0  & 0 & 0 & 0 & 0 & 0 & I\\
\mathcal{Q}_0 & \mathcal{Q}_1 & \mathcal{Q}_2 & \mathcal{Q}_3 & \mathcal{Q}_4 & \mathcal{Q}_5 & \mathcal{Q}_6 \\
\end{pmatrix}  ,
&\qquad 
\mathcal{R} = \begin{pmatrix}
I & 0 & 0 & 0 & 0 & 0 & 0 \\
0 & I & 0 & 0 & 0  & 0 & 0\\
0  & 0 & I & 0 & 0 & 0 & 0 \\
0  & 0 & 0 & I & 0  & 0 & 0\\
0  & 0 & 0 & 0 & I  & 0 & 0\\
0  & 0 & 0 & 0 & 0  & I & 0\\
0 & 0 & 0 & 0 & 0 & 0 & - \mathcal{Q}_7
\end{pmatrix}  ,
\end{align}
which can then be solved using standard routines to obtain the hybrid QNM frequencies in the sound channel.

xation time and $\etat$ is the shear viscosity. We will use the following dimensionless parameterizations of the transport coefficients: $T_2 \etat = \Cet (\et + \pt)$, and $ T_2\taut_\pi =\Ctaut$. We will choose $\Cet = 10/4 \pi$ and $\Ctaut = 5 \Cet$. This choice of transport coefficients is motivated by the fact that the MIS theory models a \textit{weakly interacting} hard sector. Note that the relation $\Ctaut = 5 \Cet $ is found in the relaxation time approximation of kinetic theory, see e.g.~\cite{Romatschke:2017ejr}. All the dynamical variables $\et$, $\vt$, $\Pit^{\mn}$ of the perturbed MIS sector will be functions of spacetime coordinates. The effective metric of the MIS sector \cref{MIS-eff-metric} will self-consistently support spacetime-dependent off-diagonal components in addition to spacetime-dependent diagonal components $\At$ and $\Bt$.

\section{Results}\label{sec:results}

\begin{figure}[ht]
   \centering
        \includegraphics[width=0.45\textwidth]{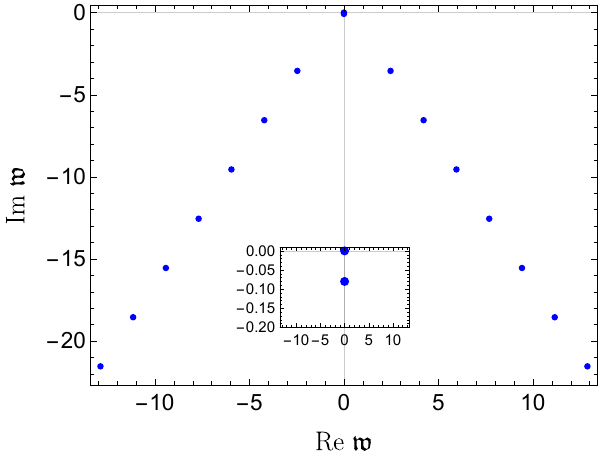}%
    \caption{
    The QNM of AdS$_4$-Schwarzschild black brane form an infinite tower of paired gapped modes arranged in the characteristic ``Christmas tree'' structure. At zero momentum, the shear hydrodynamic modes of both the holographic and MIS sectors reside at the origin. The inset zooms into the near-origin region, clearly showing the two coincident shear hydrodynamic poles at $\freq = 0$ and the MIS gapped pole lying on the negative imaginary axis.
    }
    \label{fig:christmas-tree}
\end{figure}

In this section, we describe  
the key features of the hybrid QNM in the semi-holographic model introduced earlier. First, we outline the modes present in the spectrum:
\begin{itemize}
    \item In the MIS hydrodynamic theory, there is a single shear mode in the shear channel, while the sound channel consists of a pair of sound modes. In both channels, these hydrodynamic modes\footnote{Hydrodynamic modes are low-energy gapless modes satisfying $\omega \rightarrow 0$ as $q \rightarrow 0$. Non-hydrodynamic modes are gapped modes for which $\omega \rightarrow \omega_0 \neq 0$ as $q \rightarrow 0$.} are supplemented by a single non-hydrodynamic gapped mode, which at zero momentum lies on the negative imaginary axis in the complex frequency plane. 
   \item The QNM of aAdS$_4$ black branes consist of the same hydrodynamic content as in MIS (i.e.~a single shear mode in the shear channel and a pair of sound modes in the sound channel),
   together with infinitely many pairs of non-hydrodynamic modes, which form a characteristic ``Christmas tree'' structure in the lower half of the complex frequency plane. The modes of the two decoupled sectors are plotted in \cref{fig:christmas-tree}.
    \item {In our semi-holographic model, the interplay between these two sectors gives rise to several novel and interesting features. One such feature is that two additional modes appear in the upper half of the frequency plane (UHP) due to the coupling between the sectors, both in the shear and sound channel, as observed earlier in \cite{Mondkar:2021qsf} in the context of a bulk scalar field coupled to a dynamical Goldstone mode at the boundary. {As shown in explicit numerical simulations in \cite{Mondkar:2021qsf} with scalar semi-holographic coupling, these modes in the UHP drive energy transfer from the black hole to the perturbative system at the boundary at early time (while the black hole entropy always increases), but the full non-linear dynamics lead to thermalization at longer times reaching a static final state. Furthermore, only the modes in the LHP control the long time (relaxation) dynamics.} In fact, it has been proved in \cite{Mitra:2025afp} that the global thermal equilibrium in semi-holography has maximum entropy, and typically the full system evolves towards the global thermal equilibrium at late times. We therefore ignore the modes in the UHP as these do not control long-time behavior of the full system.} 

    {Since the modes in UHP become irrelevant after a short time-scale as seen in \cite{Mondkar:2021qsf}, we expect that the modes in LHP studied here can lead to a metastable time-crystal or inhomogeneous phase at long time scales as discussed below. However, we cannot rule out the possible role of the modes in UHP in quickly destabilizing the metastable phase without explicit numerical simulation.}
\end{itemize}

In what follows, we set $\mathscr{R} =2$ and $n_2=1$. {Since both subsystems are conformal, the dispersion relations of the hybrid modes depend on $\gamma$ and $T$ only via the dimensionless combination $\gamma T^3$. We analyze the behavior of the hybrid QNM in the lower half of the complex frequency plane by varying the momentum $q$ and $\gamma T^3$.}

\subsection{Shear Channel}\label{sec:results::subsec:shear}

\paragraph{Prethermal time crystal (PTC) behavior from modified k-gap phenomenon in the hard sector:} One of our main results is that a pair of oscillating and nearly dissipationless modes emerges generically at low temperatures in the shear channel in semi-holographic systems. These modes can lead to metastable prethermal time crystal (PTC) phase as argued in the Introduction and elaborated below. These PTC modes originate from the modification of the \textit{k-gap} modes~\cite{Baggioli:2019jcm} of the hard sector described by MIS theory coupled to the holographic sector. 

The \emph{k-gap} dispersion relation for the linearized fluctuation of the decoupled MIS theory in the shear channel is  \cite{Grozdanov:2018fic, Baggioli:2019jcm}
\begin{equation}\label{eqn:k-gap}
    \freq^2 + \frac{i \freq}{\pi \Ctaut} - \frac{\Cet}{\Ctaut} \momentum^2 = 0
\end{equation}
where $\omega = \mathfrak{w} \pi T$, $q = \momentum \pi T$.
The solutions of   \cref{eqn:k-gap} are
\begin{equation}\label{eqn:k-gap:sol}
    \freq = - \frac{i}{2 \pi \Ctaut} \left( 1 \pm \sqrt{1 - 4 \Ctaut \Cet \momentum^2} \right)
\end{equation}
It is clear from the k-gap dispersion relation   \cref{eqn:k-gap:sol} that for small momenta ($4 \Ctaut \Cet \momentum^2 < 1 $), both the modes described by this dispersion relation have purely imaginary $\freq$. These are precisely the gapless hydrodynamic shear mode and the gapped relaxation mode of the MIS theory. These two modes collide with each other on the negative imaginary axis at $\freq = \freq_c :=  - \dfrac{i}{2 \pi \Ctaut} $ in the complex $\freq$ plane for $4 \Ctaut \Cet \momentum^2 = 1$. For $4 \Ctaut \Cet \momentum^2 > 1$, the frequencies of both of these modes acquire non-zero real parts which increase monotonically with $\momentum$. Additionally, for $4 \Ctaut \Cet \momentum^2 > 1$,  ${\rm Im}\,\freq$ of both the modes have a constant value $-\dfrac{1}{2 \pi \Ctaut}$.

\begin{figure}[tbp]
    \centering
        \includegraphics[width=0.5\textwidth]{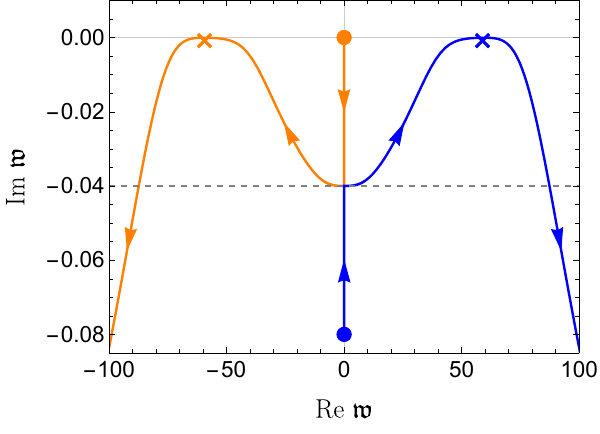} 
        \caption{\textit{Prethermal time crystal modes in the shear channel.}
The shear hydrodynamic (gapless) mode is displayed in orange and the non-hydrodynamic (gapped) mode in blue. The solid dots represent their respective positions in the complex frequency plane for $\momentum = 0$. The respective colored lines with arrows show the trajectories of these modes as $\momentum$ is increased from zero. The crosses denote the smallest $|\text{Im}(\freq)|$. We refer to these modes as prethermal time crystal (PTC) modes with the corresponding PTC frequency (crosses) and momentum denoted by $(\freq_g, \momentum_g)$. The colored lines are generated with $\gamma T^3=0.0005$, 
        while the gray dashed line denotes the trajectory of the MIS k-gap modes for $\gamma = 0$. In this figure, we have set $T=1$.
        }
        \label{fig:shear-collision}
\end{figure}

\begin{figure}[ht]
    \centering
    \subfigure[$|\text{Re}\, \freq|$ vs $\momentum$]{%
        \includegraphics[width=0.45\textwidth]{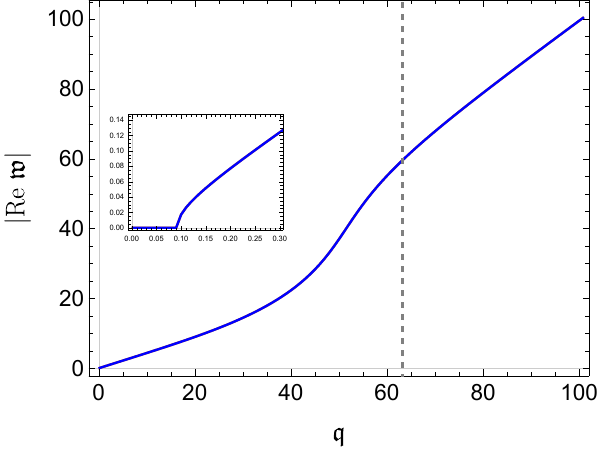}%
        \label{fig:k-gap-dispersion:sub1}}
    \hfill
    \subfigure[$\text{Im} \,\freq $ vs $\momentum$]{%
        \includegraphics[width=0.45\textwidth]{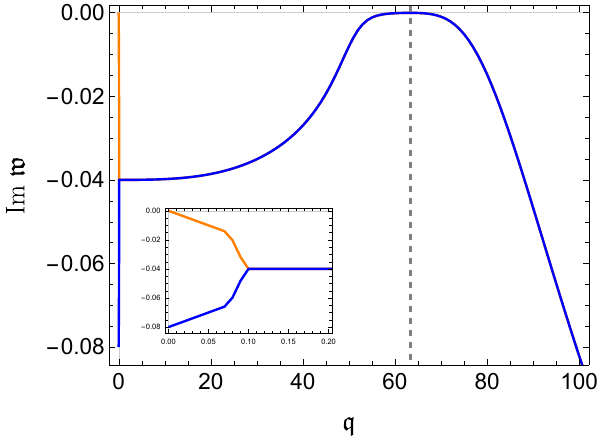}%
        \label{fig:k-gap-dispersion:sub2}}
    \\
   \caption{\textit{Prethermal time crystal modes in the shear channel: Dispersion relations.}
   Panels (a) and (b) show $|\text{Re}\, \freq (\momentum)|$ and $\text{Im} \,\freq (\momentum)$ respectively, corresponding to the PTC modes in \cref{fig:shear-collision}. All parameters are identical to those in \cref{fig:shear-collision}. The orange curve denotes the gapless branch, while the blue curve corresponds to the gapped branch. 
   $|\text{Re}\, \freq |$ is equal for both, so the two curves overlap in (a). 
   The vertical gray dashed line marks $\momentum = \momentum_g$, where $| \text{Im } \freq | $ has the samllest value. The inset panels zoom into the small-$\momentum$ region, where the dispersion relation coincides with that of the decoupled MIS sector. At larger momenta, $|{\rm Re} \freq (\momentum)|$ exhibits linear scaling with a clear crossover between two distinct slopes.
     }
    \label{fig:k-gap-dispersion}
\end{figure}

We find that the coupling to the holographic sector radically changes this behavior at higher momenta. It suppresses the imaginary part of $\freq(\momentum)$ to nearly zero at a particular momentum value, $\momentum_g$, thereby rendering the k-gap modes (almost) dissipationless. This phenomenon is absent in either sector taken alone and is a robust feature of the hybrid system at low temperatures. The trajectories of the two k-gap modes in the complex-frequency plane for a representative low value of the coupling, $\gamma T^3=0.0005$, are shown in \cref{fig:shear-collision}, while their dispersion relations are displayed in \cref{fig:k-gap-dispersion}.

In \cref{fig:shear-collision}, the blue and orange curves correspond to the gapped and gapless branches of the \textit{hybrid} k-gap dispersion relation. The solid dots mark their positions at $\momentum = 0$. As the momentum is increased from zero, the two modes first move along the negative imaginary axis in opposite directions and, as in MIS theory, collide on the negative imaginary axis at a purely imaginary frequency.
Beyond this collision, both modes acquire real parts of equal magnitude and opposite sign. Their imaginary parts, however, behave very differently from the MIS case; instead of saturating to a fixed negative value, they continue to decrease in magnitude as $\momentum$ increases. Each mode reaches a point at which $|{\rm Im}\, \freq|$ becomes minimum
indicated by the colored crosses in \cref{fig:shear-collision}—before growing again at larger momenta. We denote the corresponding momentum and frequency by $(\freq_g,\momentum_g)$. At this point, the pair of modes is almost dissipationless and propagates with equal and opposite real frequencies. These are the long-lived oscillatory excitations which we call prethermal time-crystal (PTC) modes. For $\momentum > \momentum_g$, both $|{\rm Im}\, \freq|$ and $|{\rm Re}\, \freq|$ increase 
as the modes move deeper into the complex plane. The real parts of the dispersion relations of the hybrid k-gap modes, shown in \cref{fig:k-gap-dispersion}a for $\gamma ^3= 0.0005$, also display an interesting feature. Over the entire momentum range in which the modes exist, ${\rm Re}\,\freq (\momentum)$ remains approximately linear, but with a clear crossover. At small momenta, the pair propagates with one effective velocity, while at larger momenta, the slope changes to a different value. 

\begin{figure}[tbp]
    \centering
        \includegraphics[width=0.5\textwidth]{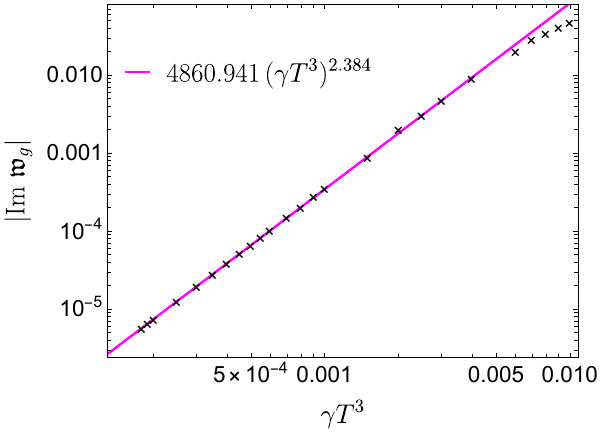} 
        \caption{The absolute value of the imaginary part of frequency, $\freq_g$, of the PTC modes of the shear channel as a function of $\gamma T^3$. The crosses are the data points. The line is the power-law fit.
        }
        \label{fig:shear-TC}
\end{figure}

\begin{figure}[ht]
    \centering
    \subfigure[$|\text{Re} \freq_g|$ vs $\gamma T^3$]{%
        \includegraphics[width=0.45\textwidth]{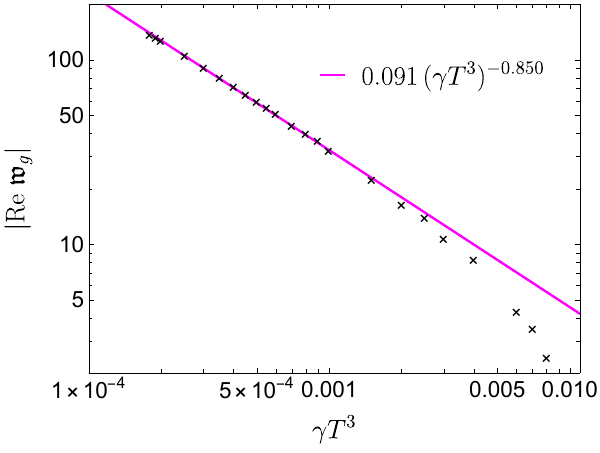}%
        \label{fig:PTC-mode-scaling-gamma-sub1}}
    \hfill
    \subfigure[$\momentum_g$ vs $\gamma T^3$]{%
        \includegraphics[width=0.45\textwidth]{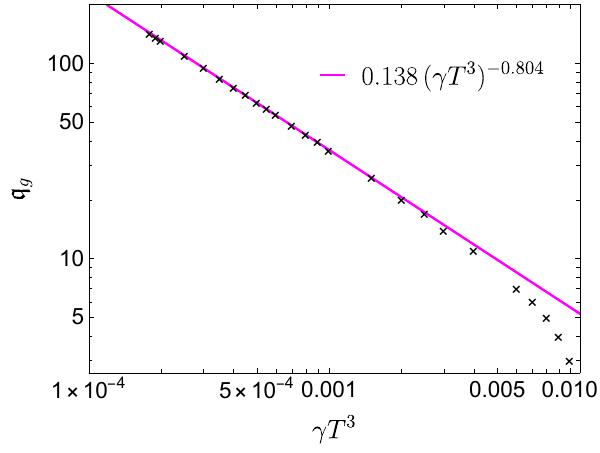}%
        \label{fig:PTC-mode-scaling-gamma-sub2}}
    \caption{(a) The absolute value of the real part of frequency, $\freq_g$, and (b) the momentum, $\momentum_g$ of the PTC modes of the shear channel as functions of $\gamma T^3$ with a power law fitting. The crosses are the data points. The lines are the power-law fits. 
    }
    \label{fig:PTC-mode-scaling-gamma}
\end{figure}

The emergence of these almost dissipationless PTC modes from the k-gap sector is a phenomenon that occurs only at sufficiently small temperatures and equivalently at small values of the coupling $\gamma$ as all physical modes depend on $\gamma$ via the dimensionless combination $\gamma T^3$. Since the lifetime of a mode is set by the inverse magnitude of the imaginary part of its frequency, we track the behavior of $|{\rm Im}\, \freq_g|$ as a function of $\gamma T^3$, shown in \cref{fig:shear-TC}. In the asymptotic limit $\gamma \rightarrow 0$, we find that $|{\rm Im} \,\freq_g|$ tends to zero, implying that the PTC modes live arbitrarily long. A power-law fit to the numerical data yields a critical exponent of $2.384$. In contrast, both $|{\rm Re} \, \freq_g|$ and $\momentum_g$ diverge as $\gamma \rightarrow 0$ ($T\rightarrow 0$), as illustrated in \cref{fig:PTC-mode-scaling-gamma}, and they do so with similar exponents of approximately $0.8$. Therefore, the observation that the PTC modes become exactly dissipationless in the $\gamma \rightarrow 0$ ($T\rightarrow 0$) limit does not contradict the fact that there are no PTC modes in the standalone MIS theory. A similar non-trivial $\gamma \rightarrow 0$ limit in collective mode behavior in semi-holography has been found earlier in \cite{Mondkar:2021qsf}.

The feature that both $|{\rm Re} \, \freq_g|$ and $\momentum_g$ diverge while $|{\rm Im} \, \freq_g|$ becomes arbitrarily small is exactly what is expected of prethermal time crystals as discussed in the Introduction. The metastable oscillating time-crystal phase is sustained by a separation of scales of the slow and fast degrees of freedom, with the oscillation period controlled by the scale of the fast system. In the semi-holographic systems, this scale separation is a result of the nature of the inter-system coupling and also emergent rather than inbuilt  (note that both subsystems in the example studied here are conformal and therefore have no intrinsic scales). For instance, when $\gamma T^3 = 0.00018$,  $\momentum_g \approx 142$, ${\rm Re} \, \freq_g \approx 138
$ and ${\rm Im} \, \freq_g \approx 5.7\times10^{-6}$.

\paragraph{Quasihydrodynamic description of PTC behavior:} Since time crystals are a collective phenomenon, it is natural to ask whether a low-energy effective theory with minimal ingredients can reproduce the PTC mode dispersion relation observed in our semi-holographic system. A promising framework to address this question is quasi-hydrodynamics~\cite{Grozdanov:2018fic}, which allows us to capture long-time, long-wavelength dynamics in the presence of softly broken continuous symmetries. Within this framework, one can construct a quasi-hydrodynamic effective theory that naturally features the PTC modes. This can be interpreted from the viewpoint of momentum relaxation \cite{Hartnoll:2016apf} with the additional feature that the usual momentum relaxation given by
\begin{align}\label{eq:quasi-conserved}
   \nabla_\mu T^\mn= \Gamma_\mu T^\mn.
\end{align}
is modified at the linear level to take the form
\begin{equation}\label{eq:quasi-conserved2}
    \nabla_\mu \delta T^\mn= \left( u^\nu u_\rho u_\sigma \tau_e^{-1}(\nabla_\perp) -  \Delta^{\nu}_{\phantom{\rho}\rho}u_\sigma\,\, (u\cdot\nabla)^2\tau_p(\nabla_\perp) \right)\delta 
    T^{\rho\sigma}.
\end{equation}
The derivative expansion of the relaxation time takes the form $\tau_e(\nabla_\perp)= \tau_{e0} + \ldots$ and $\tau_p(\nabla_\perp)= \gamma\nabla_\perp + \ldots$ In Fourier space, the linearized relation becomes 
\begin{align}
    ik_\mu \delta T^{\mn} &= 
    \begin{cases}
         \tau^{-1}_e\delta T^{00},  &\text{for } \nu=0,\\
       \omega^2\tau_p\delta T^{0i}, &\text{for } \nu=i.
    \end{cases}
\end{align}
The dimensionless momentum relaxation constant $\Ctaut$ is modified to a momentum-dependent function such that soft momentum modes are approximately conserved. This quasi-hydrodynamic theory can capture general PTC behavior, as one important condition for prethermalization is that soft momenta cannot efficiently dissipate, as discussed in the Introduction. A detailed account of this quasi-hydrodynamic construction, along with how it reproduces the PTC behavior, has been presented in \cref{app:sec:quasihydro}.

\begin{figure}[!t]
    \centering
    \subfigure[$\text{Im}\freq$ vs $\text{Re}\freq$]{%
        \includegraphics[width=0.45\textwidth]{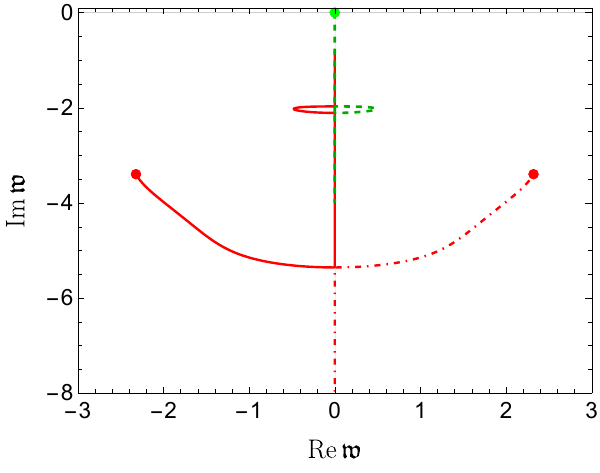}%
        \label{fig:level-crossing-sub1}}
        \vfill
    \subfigure[$\text{Re}\freq$ vs $\momentum$]{%
        \includegraphics[width=0.45\textwidth]{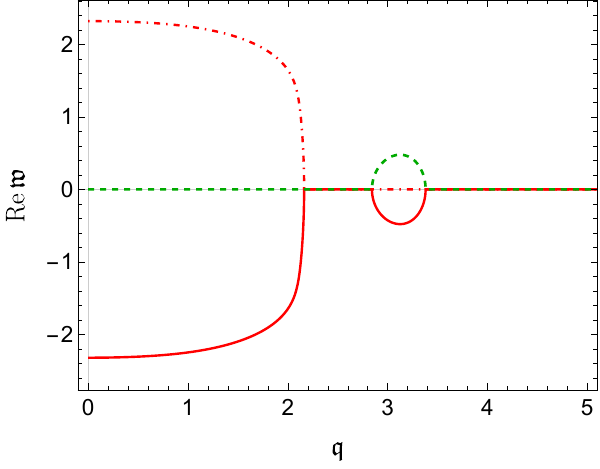}%
        \label{fig:level-crossing-sub2}}
    \hfill
    \subfigure[$\text{Im} \freq$ vs $\momentum$]{%
        \includegraphics[width=0.45\textwidth]{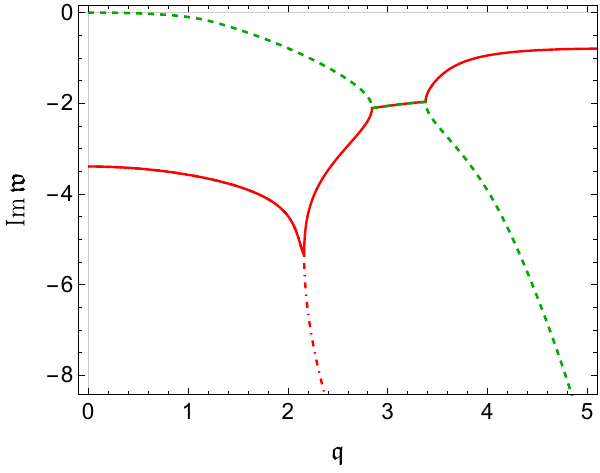}%
        \label{fig:level-crossing-sub3}}
    \caption{\textit{Level crossing in the shear channel.} (a) The gapless shear mode from the holographic sector (green) and the topmost pair of gapped QNM (red) are branches of a single dispersion relation when the sectors are strongly coupled ($\gamma =1$). The solid dots are the positions of these modes in the complex $\freq$ plane for $\momentum =0$. 
    As momentum is increased, the trajectories of these modes are shown by curves of the same colors: red solid/dot-dashed for the gapped branch, and green dashed for the gapless branch. The first collision occurs between the two red modes on the negative imaginary axis. The second collision involves the descending gapless mode and the ascending post-collision red mode, after which both modes acquire a nonzero real part and subsequently undergo a third collision back on the negative imaginary axis. Panels (b) and (c) display the corresponding dispersion relations, ${\rm Re }\,\freq (\momentum)$ and ${\rm Im }\,\freq (\momentum)$, for the same modes shown in (a).
    }
    \label{fig:level-crossing}
\end{figure}

\paragraph{\textit{Level crossing:}} Another noteworthy finding in the shear channel is the mixing of hydrodynamic and non-hydrodynamic modes at high temperatures. In the decoupled holographic sector, the shear-channel spectrum consists of a single hydrodynamic shear mode accompanied by an infinite tower of paired gapped QNM forming a Christmas tree like structure (each mode in a pair has equal real parts with opposite signs and same imaginary parts). The decoupled MIS sector, in contrast, contains one gapped mode in addition to its own shear hydrodynamic mode. When the two sectors are coupled, we find that at sufficiently high temperatures ($\gamma > \gamma_l$; $1.11 < \gamma_l < 1.12$), the hybridized holographic shear mode mixes with the uppermost pair of hybridized gapped QNM (the first branch of the Christmas tree). As a consequence, these three modes, which are originally distinct in the decoupled holographic theory, merge into interconnected branches of a single dispersion relation after coupling with the MIS sector. This behavior is illustrated in \cref{fig:level-crossing} 
for $\gamma T^3 =1$.

In \cref{fig:level-crossing}, the red solid/dot-dashed curves depict the trajectories of the top hybridized gapped QNM pair, while the dashed green curve shows the trajectory of the hybridized holographic shear mode. 
Solid dots indicate their initial positions at $\momentum = 0$. As the momentum is increased, the hybridized gapped pair moves downward and collides on the negative imaginary axis; after this first collision, one branch moves upward along the imaginary axis while the other moves downward with further increase of $\momentum$. Simultaneously, the hybridized shear mode descends from the origin along the negative imaginary axis. At a finite momentum, this descending shear mode collides with the upward-moving branch of the gapped pair—this is the key hydrodynamic–non-hydrodynamic mixing event. After this second collision, the two participating modes acquire nonzero real parts and move away from the imaginary axis before turning around and returning to it, culminating in a third collision. Following the third collision, one mode travels downward while the other moves upward along the imaginary axis; the latter eventually turns around before it could reach the origin and then descends again. The corresponding real and imaginary parts of their dispersion relations are shown in \cref{fig:level-crossing}.

Such collisions between hydrodynamic and non-hydrodynamic modes play a crucial role in determining the radius of convergence of the hydrodynamic gradient expansion~\cite{PhysRevLett.122.251601, Grozdanov:2019uhi, Gouteraux:2025kta}. This phenomenon is known as \textit{level crossing}, and the behavior we observe at high temperatures provides a concrete realization of this within the hybrid semi-holographic framework.

\subsection{Sound Channel}\label{sec:results::subsec:sound}

\begin{figure}[ht]
    \centering
    \subfigure[]{%
        \includegraphics[width=0.435\textwidth]{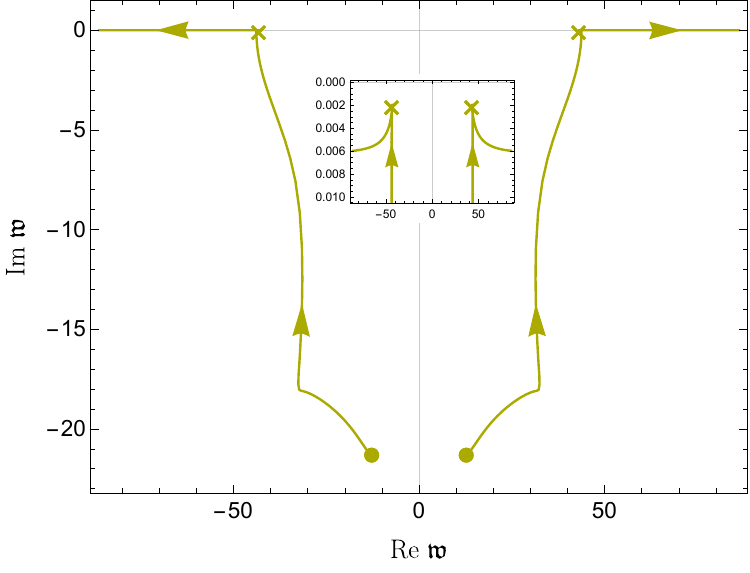}%
        \label{fig:time-crystal-sound-g=0p0008-sub1}}
        \hfill
    \subfigure[]{%
        \includegraphics[width=0.45\textwidth]{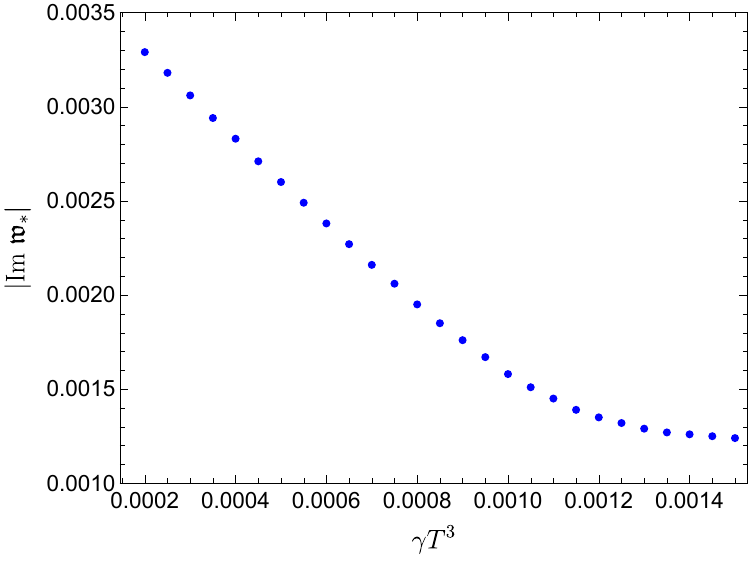}%
        \label{fig:time-crystal-sound-g=0p0008-sub2}}
    \caption{\textit{Nearly dissipationless hybrid sound-channel modes.} 
    (a) Trajectory of a representative gapped QNM pair in the sound channel (seventh from the top of the tower) for $\gamma = 0.0008$, shown in the complex $\freq$-plane. Solid dots denote the positions of the modes at $\momentum =0$, and arrows indicate the direction of flow as $\momentum$ increases. As momentum increases, the two modes approach the real axis and become almost dissipationless at $(\freq_*, \momentum_*)$, marked by crosses. The inset zooms into the near-real-axis region to highlight the long-lived pair. (b) Minimal damping $|{\rm Im} \freq_*|$ as a function of the semi-holographic coupling $\gamma T^3$, demonstrating a low-temperature window in which the sound channel hosts an oscillating, parametrically long-lived pair of hybrid modes.
    }
    \label{fig:time-crystal-sound-g=0p0008}
\end{figure}

\paragraph{Prethermal time crystal (PTC) behavior from modified QNMs of the holographic sector:} We also identify a pair of oscillating, nearly dissipationless PTC modes in the sound channel, whose origin is distinct from that of the PTC modes discussed earlier in the shear channel. In the shear channel, the PTC modes descend from the $k$-gap modes of the decoupled MIS sector. By contrast, in the sound channel, the long-lived pair originates from one of the gapped QNM pairs of the holographic sector. As the dimesionless inter-system coupling $\gamma T^3$ is varied, different hybridized gapped pairs become the longest-lived excitations: for smaller $\gamma T^3$ the relevant pair lies near the bottom of the Christmas-tree structure, whereas for larger $\gamma T^3$ it is drawn from higher up in the tower. For any fixed value of $\gamma T^3$ we find that \textit{exactly one} such hybridized pair becomes parametrically longer lived than the others.  \cref{fig:time-crystal-sound-g=0p0008} illustrates the trajectory of a representative pair
(seventh branch from the top)
that exhibits this behavior at $\gamma T^3 = 0.0008$. The solid dots mark the locations of the two modes at $\momentum = 0$, and the solid curves show their trajectories in the complex frequency plane as $\momentum$ is increased. The crosses indicate the point along each trajectory where the absolute value of the imaginary part of the frequency has the smallest value; we denote this frequency by $\freq_*$ and the corresponding momentum by $\momentum_*$.

The dependence of the minimal damping rate on temperature is summarized in \cref{fig:time-crystal-sound-g=0p0008}b. There we plot $|\text{Im}(\freq_*)|$ as a function of $\gamma T^3$. For $0.0001 \lesssim \gamma T^3 \lesssim 0.0015$ we find $|\text{Im}(\freq_*)| \lesssim 0.0035 $, with a minimum near $\gamma = 0.0015$. Thus, over a finite low-temperature window, the sound channel of the hybrid system supports an oscillating QNM pair whose damping rate is parametrically small compared to the microscopic scale set by the temperature. These modes provide a second realization of prethermal time-crystal–like behavior in our model, and they arise without any additional fine-tuning.

\paragraph{Inhomogenous zero modes:} Another noteworthy feature of the hybrid collective modes in this channel is the emergence of inhomogeneous zero modes (IZM). IZM are defined by a vanishing frequency at a finite momentum, $\freq (\momentum_0) = 0$ with $\momentum_0 \neq 0$. Recall that in the sound channel, the spectrum of decoupled MIS sector contains a pair of hydrodynamic sound modes together with a gapped mode located on the negative imaginary axis at zero momentum. The holographic sector, by contrast, exhibits a pair of hydrodynamic sound modes accompanied by an infinite tower of gapped QNM pairs. Once the two sectors are coupled, the hybridized gapped mode on the negative imaginary axis in the MIS sector continuously evolves into an IZM.

No such zero mode is present in the sound channel of the decoupled MIS theory: the MIS gapped mode asymptotes to a negative imaginary constant at large~$\momentum$ and never approaches the origin. It is only through coupling to the holographic sector that this mode is driven to $\freq=0$ at a nonzero momentum. Remarkably, we find that the resulting IZM is not restricted to low temperatures. It persists throughout the full temperature range we have explored. Moreover, as the momentum is increased beyond $\momentum_0$, this IZM shifts upward along the imaginary frequency axis into the upper half-plane, signaling a dynamical instability. {The qualitative nature of this instability is closely reminiscent of the Gregory–Laflamme (GL) instability of higher-dimensional black holes~\cite{Gregory:1993vy, Lehner:2011wc, Emparan:2015gva}, with a key distinction that the instability here is in short-wavelength modes instead of long-wavelength modes.}

{To our knowledge, this constitutes the first example of a finite-wavelength instability in an asymptotically AdS$_4$ black hole with a {planar horizon} in {pure} Einstein gravity, i.e.\ in the absence of any bulk matter fields. A related phenomenon was previously identified in~\cite{Mondkar:2021qsf}, where an IZM and a finite-wavelength instability were observed in a semi-holographic model involving a bulk scalar field. In the present case, the instability arises solely from coupling the MIS theory to the boundary degrees of freedom of the holographic sector which is described by pure Einstein's gravity.}

\begin{figure}[tbp]
    \centering
    \subfigure[]{%
        \includegraphics[width=0.45\textwidth]{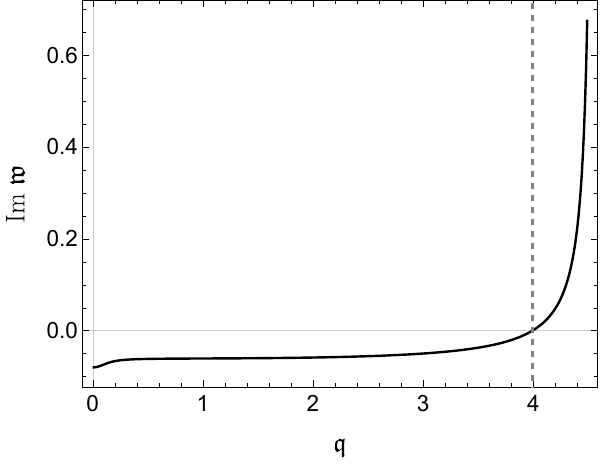}%
        \label{fig:inhomo-sound-dispersion-sub1}}
    \hfill
    \subfigure[]{%
        \includegraphics[width=0.45\textwidth]{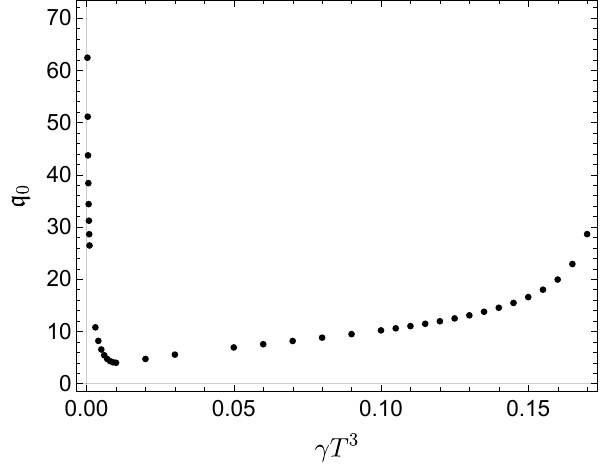}%
        \label{fig:inhomo-sound-dispersion-sub2}}
    \caption{(a) The behavior of the imaginary part of the frequency of the inhomogeneous zero mode of the sound channel with momentum for $\gamma T^3 = 0.01$. As $\momentum$ is increased, this mode moves up the negative imaginary axis, crosses the origin at $\momentum_0$, and subsequently becomes 
    unstable. The vertical dashed line marks $\momentum = \momentum_0$. (b) The dependence of the zero mode momentum, $\momentum_0$, on the semi-holographic coupling. 
    }
    \label{fig:inhomo-sound-dispersion}
\end{figure}

The left panel of  \cref{fig:inhomo-sound-dispersion} illustrates the imaginary part of the frequency as a function of $\momentum$ for a representative value $\gamma=0.01$, showing the emergence of the IZM and its subsequent motion into the upper half-plane. The right panel of  \cref{fig:inhomo-sound-dispersion} displays the dependence of the zero-mode momentum $\momentum_0$ on~$\gamma$ (or equivalently, on temperature). We observe that $\momentum_0\to \infty$ as $\gamma\to 0$. As $\gamma$ increases, $\momentum_0$ initially decreases, reaches a minimum near $\gamma\simeq 0.01$, and then grows with further increase in~$\gamma$, demonstrating that the IZM persists even at higher temperatures. For $\gamma\gtrsim 0.17$, $\momentum_0$ grows very rapidly, making its precise numerical extraction increasingly challenging with our code. This temperature dependence contrasts with the semi-holographic model of~\cite{Mondkar:2021qsf}, where the IZM appeared only at low temperatures.

Physically, an inhomogeneous zero mode is the signal that the homogeneous thermal state wants to develop structure: a static fluctuation with non-zero wavelength that neither decays nor grows. In our model, this happens in the longitudinal (sound) channel, where the hybridized gapped MIS mode crosses the origin at a finite momentum $\momentum_0$. For $\momentum < \momentum_0$  the mode has ${\rm Im}\, \freq < 0$ and inhomogeneous perturbations relax back to equilibrium, while for $\momentum > \momentum_0$ the same branch moves into the upper half–plane and becomes unstable. The point $\freq(\momentum_0) = 0$ therefore marks the onset of a short wavelength instability of the planar black brane, with a preferred length scale $2 \pi/\momentum_0$. From the quasihydrodynamic viewpoint spelled out in \cref{app:sec:quasihydro}, this mode can be interpreted as an energy-density fluctuation whose relaxation time $\tau_e(\momentum)$ becomes momentum-dependent and precisely cancels the sound-mode damping at $\momentum=\momentum_0$, converting a purely damped non-hydrodynamic excitation into a static deformation of the energy density.

{Finite-momentum zero modes also appear in Fulde–Ferrell–Larkin–Ovchinnikov states and in holographic spatially modulated phases, where they usually signal a finite-wavevector instability of a homogeneous phase toward a translation-breaking equilibrium state with spatially modulated condensates or charge/current densities~\cite{RevModPhys.76.263, Donos:2011qt, Alsup:2012kr, Withers:2013kva}. In that sense, the similarity with our IZM is the emergence of a preferred nonzero wavevector and an associated length scale. The crucial difference is that the instability in those works is typically an order-parameter instability tied to the onset of a new thermodynamic phase, whereas in our case the hybrid non-hydrodynamic mode is damped for $\momentum < \momentum_0$ and becomes unstable for $\momentum > \momentum_0$, so the finite-$\momentum$ zero mode marks the threshold of a genuinely dynamical short-wavelength instability.}

\FloatBarrier

\section{Discussion}\label{sec:discussion}

Our results uncover several qualitatively new dynamical features of semi-holographic systems and planar AdS$_4$ black holes governed by pure Einstein gravity with dynamical boundary conditions. The first important result concerns the emergence of prethermal time-crystal modes in the shear and sound channels in the hard and soft sectors, respectively. The MIS k-gap modes, when softly coupled to the holographic sector, develop an extended regime in which their imaginary parts are strongly suppressed, producing long-lived oscillatory excitations that closely resemble prethermal time crystals. These oscillatory modes arise without any fine-tuning, purely from the structure of the semi-holographic coupling and are absent in either sector on its own. Similarly, the holographic quasi-normal modes in the sound channel obtained from black brane perturbations are modified by the coupling to the hard sector to develop almost dissipationless behavior at finite frequency and momentum. The appearance of these phenomena at low temperatures demonstrates that hybrid weak–strong systems can exhibit robust time-periodic dynamics even without external driving.

The second key finding is that the inhomogeneous zero mode identified previously in \cite{Mondkar:2021qsf} in the context of spontaneous symmetry breaking in semi-holographic systems, which appeared only at low temperatures, persists at all temperatures in the present model. {To our knowledge, this represents the first example of a finite-wavelenth instability of a planar AdS$_4$ black brane in two-derivative pure Einstein gravity triggered entirely by boundary interactions, without invoking bulk matter fields or higher-derivative corrections.}

Our analysis also reveals a concrete instance of hydrodynamic–nonhydrodynamic mixing in a hybrid system. At sufficiently high temperatures, the shear hydrodynamic mode of the holographic sector and the topmost gapped quasi-normal-mode pair collide and reorganize into branches of a single dispersion relation. This is directly analogous to the pole collisions in purely holographic plasmas and Müller–Israel–Stewart theories that control the analytic structure and convergence of the hydrodynamic gradient expansion, as established in Refs.~\cite{PhysRevLett.122.251601, Grozdanov:2019uhi, Gouteraux:2025kta}. Our results extend this phenomenon to genuinely semi-holographic systems: here, the colliding modes are already hybrid objects built from both sectors, and their reorganization occurs alongside the appearance of prethermal time-crystal behavior and inhomogeneous zero modes. Taken together, these results demonstrate that semi-holography provides a controlled framework in which planar black holes can exhibit prethermal oscillatory phases, robust finite-momentum instabilities, and hybrid level crossings—all within pure Einstein gravity. They also show that coupling even simple hydrodynamic degrees of freedom to a strongly interacting holographic plasma suffices to generate rich dynamical behavior unavailable in either sector alone. 
We emphasize that it would be necessary to solve the non-linear dynamics of the full system involving numerical relativity coupled to the dynamical MIS sector at the boundary to establish the novel dynamical phases firmly.

It is worth mentioning that long-lived oscillations in the presence of chiral anomaly have also been observed in holographic systems 
with external homogeneous magnetic fields and perpendicular time-dependent electric fields \cite{Ammon:2016fru,Haack:2018ztx,Grieninger:2023wuq} (see also \cite{Waeber:2024ilt}). However, this does not constitute time-crystal-like behavior as the external driving is a quench (has no discrete symmetry). More recently, spontaneous breaking of the discrete time symmetry of periodic driving in holographic systems has been reported in \cite{Yang:2023dvk,Tai:2025gvw}. As an outlook, it would be interesting to understand the generic circumstances in which both discrete and continuous time translation symmetry can be spontaneously broken in holographic and semi-holographic systems.

\begin{acknowledgments}

We thank Himadri Dhar, Sa{\v s}o Grozdanov,  Giuseppe Policastro, Anton Rebhan, Saptarshi Saha and Ujjwal Sen for helpful discussions. TM has been supported by the DFG through the Emmy Noether Programme (project
number 496831614), through CRC 1225 ISOQUANT (project number 27381115). The research of AM has been supported by
the FONDECYT regular grant no. 1240955 of La Agencia Nacional de Investigación y Desarrollo (ANID), Chile. AS is supported by funding from Horizon Europe research and innovation programme under the Marie Sk\l odowska-Curie grant agreement No.~101103006 and the project N1-0245 of Slovenian Research Agency (ARIS).
\end{acknowledgments}

\appendix

\section{Near-boundary expansions}\label{app:near-boundary-exp}
The near-boundary expansions of the bulk metric perturbations can be obtained by solving linearized bulk Einstein equations order-by-order near the spacetime boundary ($r \to \infty$). The leading order terms in this expansion are the respective components of the perturbed effective metric of the soft sector. Here we provide key details in both the shear and sound channel computations.

\subsection{Shear channel}\label{app:near-boundary-exp::subsec-shear}

Recall that in the shear channel, the components $ \D \Gtx(r) $ and $\D \Gyx (r)$ of the bulk metric $G_{MN}$ are perturbed. 
The near-boundary expansions of the bulk metric perturbations $\D G_{\mn}(r)$ takes the general form
\begin{align}
    \D G_{\mn} = \sum_{i=0}^{\infty} r^{2-i} \D g^{(i)}_{\mn}
\end{align}
The leading term in this expansion, namely $\D g^{(0)}_{\mn}$, is the non-normalizable mode which is identified with $\D g_{\mn}$, i.e., with $\mn$- component of the perturbed effective metric of the soft sector. The subleading term in $3+1$ dimensions is $\D g^{(3)}_{\mn}$, which is the normalizable mode that cannot be determined solely from the near-boundary expansion. One needs to solve the full linearized bulk equations of motion subject to regularity at the horizon to determine the normalizable mode.

Their near-boundary expansions read 
\begin{align}
    \D \Gtx(r) &= \D \gtx r^2 - \frac{q \left( \omega  \D \gyx + q \D \gtx\right) }{2 b^2} r + 
    \D \gtxn + \frac{3 \mathrm{i} a q}{4 b^2}  \frac{\D \gyxn}{r}  + \mathcal{O}\left( \frac{1}{r^2} \right) \\
    \D \Gyx (r) &= \D \gyx r^2 - \frac{\mathrm{i}\left( \omega  \D \gyx + q \D \gtx\right)  }{a } r + \D \gyxn  \nonumber \\
    &+ \frac{ \left( a q^4 - 6 \mathrm{i} b^4 \Mbh \omega \right) \left( \omega  \D \gyx + q \D \gtx \right) + 6 \mathrm{i} \left(a^2 b^2 q^2 - 4 b^4 \omega^2 \right)}{24 a b^4 \omega} \frac{\D \gyxn}{r} + \mathcal{O}\left( \frac{1}{r^2} \right)
\end{align}
where $\D \gtxn$ and  $\gyxn$ are the normalizable modes of $\D \Gtx(r)$ and $\D \Gyx(r)$, respectively. $\D \gtxn$ is not independent but it is related to $\D \gyxn$ by the relation
\begin{equation}
    \D \gtxn =  \frac{ a q \left( \mathrm{i} q^2 \left( \omega \D \gyx  +   q \D \gtx \right) - 6 a b^2  \D \gyxn \right)}{6 b^4 \omega}
\end{equation}
The near-boundary expansion for the gauge-invariant perturbation $\Zsh(r)$ defined in \cref{eq:Zsh} then becomes
\begin{align}\label{eq:Zsh-near-boundary-exp}
    Z_{sh}(r) = \left(\omega \D \gyx + q \D \gtx \right)r^2 - \frac{\mathrm{i} \omega}{a} \left(\omega \D \gyx + q \D \gtx \right) r - \frac{q^2}{2 b^2} \left(\omega \D \gyx + q \D \gtx \right)  +  \frac{\zshn}{r} + \mathcal{O}\left( \frac{1}{r^2} \right)
\end{align}
where the normalizable mode $\zshn$ can be expressed in terms of normalizable modes $\D \gyxn$ and $ \D \gtxn$ via
\begin{equation}\label{eq:zsh3}
    \zshn = \omega \D \gyxn + q \D \gtxn
\end{equation}

\subsection{Sound channel}

The perturbed bulk metric components in the sound channel are, namely, $\D \Gtt(r)$, $\D \Gxx (r)$, $ \D \Gyy(r)$, and $\D \Gty (r)$. Their near-boundary expansions are
\begin{align} 
    \D \Gtt (r) &= \D \gtt r^2 + \D \gttres r - \left(  \frac{ \D \gxx}{2 b^4} + \frac{\imag \D \gttres \omega }{a}  - \frac{q^2 \D \gtt + 2 \omega q \D \gty + \left(\D \gxx + \D \gyy \right) \omega^2}{2 b^2}\right)  \nonumber \\
    &+ \frac{\D \gttn}{r} + \mathcal{O}\left( \frac{1}{r^2} \right) \\
    \D \Gxx (r) &= \D \gxx r^2 - \left( \frac{b^2 \D \gttres + \imag a \omega \D \gxx }{a^2} \right) r 
    + \frac{\D \gxxn}{r}  + \mathcal{O}\left( \frac{1}{r^2} \right) \\
    \D \Gyy(r)  &= \D \gyy r^2 - \left(  \frac{b^2 \D \gttres + \imag a \left( 2 q \D \gty + \omega \D \gyy \right)}{a^2} \right) r  
     -\frac{\D \gyyn}{r}  + \mathcal{O}\left( \frac{1}{r^2} \right) \\
    \D \Gty (r) &= \D \gty r^2  - \frac{\imag q \D \gtt}{2 a}  r + \frac{q}{2}\left( - \frac{\imag \D \gttres }{a} + \frac{\omega \D \gxx}{b^2} \right)  
    +  \frac{\D \gtyn}{r}  + \mathcal{O}\left( \frac{1}{r^2} \right)
\end{align}
The sub-leading term $\D \gttres$ above arises because of the residual gauge freedom in the perturbed bulk metric in the sound channel. Out of the four normalizable modes $\D \gttn$, $\D \gxxn$, $\D \gyyn$, $\D \gtyn$, only one is independent. We choose the independent one to be $\D \gttn$. The other three can be expressed in terms of it as follows
\begin{align}
    \D \gxxn &= \frac{1}{6 a^4 b^2 q^2}\Big(  \imag a^3 q^4 \omega \D \gxx - 4 b^6 \Mbh \omega^2 \D \gtt
    + \imag a b^2 q^2 \omega \left( q^2 \D \gtt + 2 \omega q \D \gty + \left(\D \gyy - \D \gxx \right) \omega^2  \right)  \nonumber \\
    &- a^2 b^4 \Mbh \left( q^2 \D \gtt + 6 \omega q \D \gty + 3 \left(\D \gyy + \D \gxx \right) \omega^2  \right) - \left(-2 a^2 b^4 q^2 + 4 b^6 \omega^2 \right) \D \gttn \Big) \\
    \D \gyyn &= - \D \gxxn \\
    \D \gtyn  &= \frac{1}{6 a^2 b^4 q} \Big( \imag a^3 q^4 \D \gxx  - 4 b^6 \Mbh \omega \D \gtt   - 3 a^2 b^4 \Mbh \left( 2 q \D \gty + (\D \gxx + \D gyy) \omega \right) \nonumber \\
    &+ \imag a b^2 q^2 \left(  q^2 \D \gtt + 2 \omega q \D \gty + (\D \gyy - \D \gxx) \omega^2 \right) - 4 b^6 \omega \D \gttn \Big)
 \end{align}
The near-boundary expansion for the gauge-invariant perturbation $\Zso(r)$ defined in  \cref{eq:Zso} then becomes
 \begin{align}\label{eq:Zso-near-boundary-exp}
    \Zso(r) &=  \left( \omega^2 \left( \D \gyy - \D \gxx \right) + 2 \omega  q \D \gty + q^2 \left(\D \gtt +\frac{a^2 \D \gxx}{b^2} \right)  \right) r^2 \nonumber \\
    &- \frac{\imag \omega }{a}  \left( \omega^2 \left( \D \gyy - \D \gxx \right) + 2 \omega  q \D \gty + q^2 \left(\D \gtt +\frac{a^2 \D \gxx}{b^2} \right)  \right)  r  \nonumber \\
    & - \frac{q^2}{2 b^2}  \left( \omega^2 \left( \D \gyy - \D \gxx \right) + 2 \omega  q \D \gty + q^2 \left(\D \gtt +\frac{a^2 \D \gxx}{b^2} \right)  \right)  + \frac{\zson}{r} + \mathcal{O}\left( \frac{1}{r^2}\right)
\end{align}
where the normalizable mode $\zson$ can be expressed in terms of normalizable modes $\D \gttn$, $\D \gxxn$, $\D \gyyn$ $ \D \gtxn$ as
\begin{align}\label{eq:zso3}
    \zson = q^2 \D \gttn + 2 \omega q \D \gtyn + \frac{a^2 q^2 \Mbh \D \gxx}{2 b^2} +  \frac{a^2 q^2 }{b^2}  \D \gxxn + \left(- \D \gxxn + \D \gyyn \right)  \omega^2 
\end{align}
Note crucially that the gauge-invariant perturbation $\Zso(r)$ is independent of the gauge-dependent term $\D \gttres$.

\section{Quasihydrodynamics}\label{app:sec:quasihydro}

We modify the conservation equations by adding a small source as in  \cref{eq:quasi-conserved}.

If $\Gamma^0=0$, then we just have weak relaxation of the shear momentum flux, i.e. $\Gamma^\mu = (0,\Gamma, \Gamma, 
0)$. In principle, $\Gamma$ (and its inverse, the relaxation time, $\tau$) can be an arbitrary function of frequency and momentum. One can think of the non-conservation as arising from a dilute weak gas whose dynamics weakly influence the fluid, see e.g.~the discussion in Chapter 5 of \cite{Hartnoll:2016apf} and also for kinetic theory computations \cite{Amoretti:2023hpb,Bajec:2024jez}.

To reproduce the results in the main body of the text from a purely (quasi-)hydrodynamics perspective, we promote $\Gamma_\mu$ to be a function of frequency and momentum, i.e.~$\Gamma=\Gamma(\omega,k)=\tau_p(k) \omega^2$. The factor of $\omega^2$ is chosen explicitly so as not to spoil the two-pole structure in the dispersion relation. 

The relevant equation arising from the conservation equation,  \cref{eq:quasi-conserved}, is the shear mode
\begin{align}
    -i\omega \delta T^{0i}+ik_j \delta T^{ij}=\tau_p(k)\omega^2 \delta T^{0i}.
\end{align}
Expanding to lowest order in gradients, we can write $\delta T^{ij}=
- \eta \sigma^{ij}= -2i D k^{<j}\delta T^{i>0}$, 
where $\eta$ is the shear viscosity, $\sigma^{ij}$ is the shear tensor, $D=\eta/(\varepsilon+P)$ is the shear diffusion constant and the angular brackets denote the transverse traceless components. 

Computing the modes of the above leads to the following dispersion relation
\begin{align}
    \omega_\pm =   \frac{1 \pm\sqrt{1-4 D k^2 \tau_p}}{2i \tau_p}.
\end{align}
This results in a theory with both a gapped and gapless mode, i.e.~for small $k$
\begin{align}
    \omega_i=-i D k^2 +\mathcal{O}(k^3), \quad \omega_+= -\frac{i}{\tau_p}+i D k^2 +\mathcal{O}(k^3).
\end{align}
Note that $\tau$ and, in principle, the shear diffusion coefficient admit a gradient expansion in $k.$ 

\begin{figure}[h]
    \centering
    \subfigure[]{\includegraphics[width=0.45\linewidth]{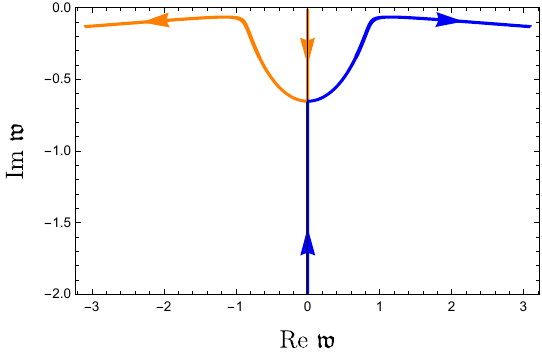}}
    \hspace{4mm}
    \subfigure[]{\includegraphics[width=0.45\linewidth]{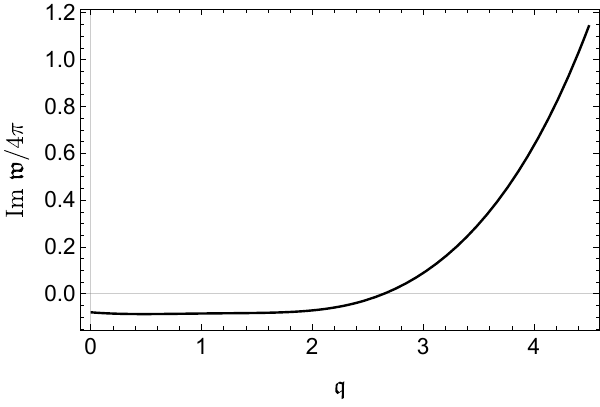}}
    \caption{(a) Shear modes in the case of weakly broken momentum conservation on a timescale $\tau(k)$ given by  \cref{eq:tau} with $\pi T=1$. Arrows indicate increasing momentum. (b) Inhomogeneous mode in the longitudinal sector with weakly broken energy conservation, given by  \cref{eq:tau-e}. The normalization is chosen for easier comparison to  \cref{fig:inhomo-sound-dispersion}.}
    \label{fig:quasihydro}
\end{figure}

As an indicative example, we show in \cref{fig:quasihydro} the behavior of the modes
where 
\begin{align}\label{eq:tau}
    \tau=\frac{c k}{1+a k +b k^2},
\end{align}
for $a=-\frac{1}{2}$, $b=0.1$ and $c=1.$ Furthermore, we take the diffusion coefficient to be a constant, $D=1.$ As in \cref{fig:shear-collision}, the collision occurs for some value of $k=k_*,$ after which the modes lose almost all of their imaginary value for some $k_g$.

In the longitudinal sector (with momentum aligned along the velocity perturbation), we are interested in the non-conservation of the energy equation in  \cref{eq:quasi-conserved} with $\Gamma_0 \neq 0$, i.e.
\begin{align}
    -i\omega \delta T^{00}+ik_i \delta T^{0i}=\tau_e^{-1} \delta T^{00},
\end{align}
where we introduced the relaxation time for the energy density.
The gapped nonhydrodynamic mode reads to lowest order in $k$
\begin{align}
    \omega= - \frac{i}{\tau_e}+ i \tau_e c_s^2 k^2.
\end{align}
We can make contact with the fitting function  \cref{eq:tau} above, but we note that in the limit of vanishing momenta, the nonhydrodynamic mode remains gapped. The simplest option, keeping the mode gapped, is choosing 
\begin{align}\tau_e=c k/\tau(k),\label{eq:tau-e}\end{align} which we show in   \cref{fig:quasihydro}, which qualitatively agrees with the behavior in  \cref{fig:inhomo-sound-dispersion}.

\bibliographystyle{jhep}
\bibliography{main}


\end{document}